\documentclass[numberedappendix, iop, apj, revtex4]{emulateapj}
\usepackage{natbib}
\usepackage{url}
\usepackage{amsmath} 

\usepackage{appendix}
\usepackage{hyperref} 

\begin{document}

\title{UVUDF: Ultraviolet Through Near-infrared Catalog and Photometric Redshifts of Galaxies in the Hubble Ultra Deep Field}

\author{
  Marc Rafelski\altaffilmark{1,2,3}, 
  Harry I. Teplitz\altaffilmark{3}, 
  Jonathan P. Gardner\altaffilmark{1},
  Dan Coe\altaffilmark{4}, 
  Nicholas A. Bond\altaffilmark{1},  
  Anton M. Koekemoer\altaffilmark{4}, 
  Norman Grogin\altaffilmark{4},
  Peter Kurczynski\altaffilmark{5}, 
  Elizabeth J. McGrath\altaffilmark{6},
  Matthew Bourque\altaffilmark{4},
  Hakim Atek\altaffilmark{7},
  Thomas M. Brown\altaffilmark{4}, 
  James W. Colbert\altaffilmark{3}, 
  Alex Codoreanu\altaffilmark{8}, 
  Henry C. Ferguson\altaffilmark{4},
  Steven L. Finkelstein\altaffilmark{9}, 
  Eric Gawiser\altaffilmark{5}, 
  Mauro Giavalisco\altaffilmark{10},
  Caryl Gronwall\altaffilmark{11,12}, 
  Daniel J. Hanish\altaffilmark{3},
  Kyoung-Soo Lee\altaffilmark{13}, 
  Vihang Mehta\altaffilmark{14}, 
  Duilia F. de  Mello\altaffilmark{15,1},
  Swara Ravindranath\altaffilmark{4}, 
  Russell E. Ryan\altaffilmark{4},
  Claudia Scarlata\altaffilmark{14},
  Brian Siana\altaffilmark{16},
  Emmaris Soto\altaffilmark{15},
  Elysse N. Voyer\altaffilmark{17}
}

\altaffiltext{1} {Goddard Space Flight Center, Code 665, Greenbelt, MD 20771, USA marc.a.rafelski@nasa.gov}
\altaffiltext{2}{NASA Postdoctoral Program Fellow}
\altaffiltext{3}{Infrared Processing and Analysis Center, MS 100-22, Caltech, Pasadena, CA 91125}
\altaffiltext{4}{Space Telescope Science Institute, 3700 San Martin Drive Baltimore, MD 21218}
\altaffiltext{5}{Department of Physics and Astronomy, Rutgers University, Piscataway, NJ 08854}
\altaffiltext{6}{Department of Physics and Astronomy, Colby College, Waterville ME, 04901, USA}
\altaffiltext{7}{Laboratoire d'Astrophysique, \'{E}cole Polytechnique F\'{e}d\'{e}rale de Lausanne (EPFL), Observatoire, CH-1290, Switzerland}
\altaffiltext{8}{Centre for Astrophysics and Supercomputing, Swinburne University of Technology, Hawthorn, Victoria 3122, Australia}
\altaffiltext{9}{Department of Astronomy, The University of Texas at Austin, Austin, TX 78712}
\altaffiltext{10}{Astronomy Department, University of Massachusetts, Amherst, MA 01003}
\altaffiltext{11}{Department of Astronomy \& Astrophysics, The Pennsylvania State University, University Park, PA, 16802}
\altaffiltext{12}{Institute for Gravitation and the Cosmos, The Pennsylvania State University, University Park, PA 16802}
\altaffiltext{13}{Department of Physics, Purdue University, 525 Northwestern Avenue, West Lafayette}
\altaffiltext{14}{Minnesota Institute for Astrophysics, School of Physics and Astronomy, University of Minnesota, Minneapolis, MN 55455}
\altaffiltext{15}{Department of Physics, The Catholic University of America, Washington, DC 20064}
\altaffiltext{16}{Department of Physics \& Astronomy, University of California, Riverside, CA 92521}
\altaffiltext{17}{Aix Marseille Universit\'{e}, CNRS, LAM (Laboratoire d'Astrophysique de Marseille) UMR 7326, 13388, France}

\begin{abstract}

We present photometry and derived redshifts from up to eleven bandpasses for 9927 galaxies in the Hubble Ultra Deep field (UDF), covering an observed wavelength range from the near-ultraviolet (NUV) to the near-infrared (NIR) with Hubble Space Telescope observations. Our Wide Field Camera 3 (WFC3)/UV F225W, F275W, and F336W image mosaics from the ultra-violet UDF (UVUDF) imaging campaign are newly calibrated to correct for charge transfer inefficiency, and use new dark calibrations to minimize background gradients and pattern noise. Our NIR WFC3/IR image mosaics combine the imaging from the UDF09 and UDF12 campaigns with CANDELS data to provide NIR coverage for the entire UDF field of view. We use aperture-matched point-spread function corrected photometry to measure photometric redshifts in the UDF, sampling both the Lyman break and Balmer break of galaxies at $z\sim0.8-3.4$, and one of the breaks over the rest of the redshift range. Our comparison of these results with a compilation of robust spectroscopic redshifts shows an improvement in the galaxy photometric redshifts by a factor of two in scatter and a factor three in outlier fraction over previous UDF catalogs. The inclusion of the new NUV data is responsible for a factor of two decrease in the outlier fraction compared to redshifts determined from only the optical and NIR data, and improves the scatter at $z<0.5$ and at $z>2$. The panchromatic coverage of the UDF from the NUV through the NIR yields robust photometric redshifts of the UDF, with the lowest outlier fraction available.
\end{abstract}


\nocite{*} 

\section{Introduction}
\label{intro}

The Hubble Ultra Deep Field \citep[UDF;][]{Beckwith:2006} is one of the most studied fields on the sky, with extremely sensitive high-resolution imaging covering many photometric bandpasses. While only covering 12 arcmin$^2$ in the sky, the data's depth, resolution, and wavelength coverage enable a wide range of scientific work. The data have been used to measure the colors and luminosity function of high redshift galaxies \citep[e.g.,][]{Bouwens:2006, Ryan:2007, Cameron:2009, Bouwens:2011b, Dunlop:2013, Schenker:2013, McLure:2013, Oesch:2013, Bouwens:2014a}, to observe the evolution in the star formation rate density \citep[e.g.,][]{Bouwens:2007, Bouwens:2009ik, Bouwens:2010}, to determine the morphological properties of galaxies \citep[e.g.,][]{Bouwens:2004, Elmegreen:2005d, Straughn:2006,Hathi:2008, Elmegreen:2010, Oesch:2010b, Elmegreen:2014}, to constrain the star formation efficiency of gas at $z\sim3$ \citep{Wolfe:2006, Rafelski:2011}, to characterize new types of galaxies, such as clumpy galaxies \citep{Elmegreen:2007, Guo:2012}, and for many other investigations.

The upgrade of the Hubble Space Telescope (HST) in the fourth servicing mission added the Wide Field Camera 3 (WFC3) to HST's instrumentation, enabling very deep near-infrared (NIR) \citep[UDF09 and UDF12;][]{Oesch:2010a,Oesch:2010b,Bouwens:2011b, Koekemoer:2013, Ellis:2013} and near-ultraviolet (NUV) \citep[UVUDF;][]{Teplitz:2013} observations of the UDF. The NIR data enable the study of galaxies at the highest redshifts at $z>7$ \citep[e.g.,][]{Oesch:2010a, Oesch:2010b, Bouwens:2010, Finkelstein:2010, Bouwens:2011b, Bouwens:2013cz, Ellis:2013,Dunlop:2013}, while the NUV observations enable studies of the rest-frame ultraviolet (UV) at intermediate redshifts, $z\sim1$ \citep{Teplitz:2013, Bond:2014, Kurczynski:2014,Mei:2014vg}. 

Reliable redshifts of galaxies in the UDF are needed for such studies, but only a small number of galaxies in the UDF have spectroscopic redshifts, because of the faintness of the galaxies sampled (see Section \ref{specz}). Therefore, most studies in the UDF rely on  color selection techniques or photometric redshifts, or use a small number of galaxies. While color selection techniques are very useful for selecting a specific type of galaxy in a redshift interval \citep[e.g.,][]{Steidel:2003}, photometric redshifts have the advantage of determining redshifts for a large sample of galaxies, making use of all the photometric bandpasses simultaneously, while providing uncertainties on the resultant redshifts \citep[e.g.,][]{Benitez:2000}. The only public photometric redshift catalog covering the entire UDF is presented in \citet{Coe:2006}, and was created before the new NIR and NUV observations of the UDF. Since then, other redshift catalogs for subsets of the data have been released, but cover fractions of the field of view (FOV) and smaller numbers of galaxies \citep[e.g.,][]{Rafelski:2009,Cameron:2011}. 

Photometric redshifts are best determined when including strong features in the spectral energy distributions (SEDs), and galaxies exhibit multiple such features enabling robust redshift determinations. The most apparent feature is the Lyman break, composed of the Lyman limit at 912\AA~ and the Lyman series lines shortward of 1216\AA. This is followed by the Balmer break at 3646\AA, and then the 4000\AA~ break, which is composed of the CaII H and K lines and the sudden onset of photospheric opacity by ionized metals \citep{Hamilton:1985}.

This paper presents photometric redshifts that include high resolution NUV and NIR data, complementing the optical data and thereby covering the wavelength range $0.2 \lesssim \lambda \lesssim 1.8\micron$, providing the best dataset to measure redshifts in the UDF. The data enable measurements of both the Lyman break and the Balmer breaks simultaneously in the interval $0.8 \lesssim z \lesssim3.4$. Additionally, at least one of these breaks, along with a long baseline measurement of the SED slope, is observed over the entire redshift range. This effectively removes the redshift degeneracies observed when limited to optical data, and thereby significantly reduces the outlier fraction when comparing photometric redshifts to a spectroscopic redshift sample \citep[e.g.,][]{Coe:2006, Ilbert:2006, Rafelski:2009}. 

These are the first photometric redshifts that make use of the new WFC3 NUV data, which improve the redshifts by sampling the Lyman break of high redshift galaxies, and more clearly defines the 4000\AA ~break for low redshift galaxies. The NUV photometry requires careful calibration to obtain robust photometry, due to challenges with charge transfer efficiency and dark calibrations. The methods to overcome these challenges and produce the high quality NUV image mosaics, which are now available on MAST,\footnote{\url{http://archive.stsci.edu/prepds/uvudf/}} are described, and the NUV photometry is provided.

In addition to the NUV data, we include NIR data covering the entire UDF. Previous work included the deep NIR data from the UDF09 and UDF12 campaigns; in this paper the photometric redshifts make use of both of these deep NIR data and shallower NIR observations covering the entire UDF FOV from CANDELS as described below. The NIR data measures the 4000\AA~break at intermediate redshifts and provides more photometric points to constrain the SED slope, thereby improving the photometric redshifts.

The inclusion of these two new datasets make the redshifts estimated here the most accurate and robust redshifts available at this time, with significant improvements for most of the ten thousand galaxies in the UDF. The photometry and redshifts tabulated here have already been used by various studies \citep{Bond:2014, Kurczynski:2014, Mei:2014vg}, and are now available to the public.

The paper is organized as follows: In Section \ref{obs}, we describe the observations, we explain the new calibrations of the NUV data, and we characterize the data. In Section \ref{photometry}, we detail the methodology for aperture-matched PSF photometry and catalog definition. In Section \ref{photoz} we determine the photometric redshifts, and assess their quality and improvements with the addition of the NUV data. We describe the final photometric and redshift catalog in Section \ref{catalog}, and summarize the paper in Section \ref{summary}. The Appendix includes a careful description of the need for specialized dark calibrations for WFC3/UVIS and the methodology used to implement them.

\section{Observations}
\label{obs}

\begin{figure}[t!]
\center{
\includegraphics[scale=0.4, viewport=10 10 550 550,clip]{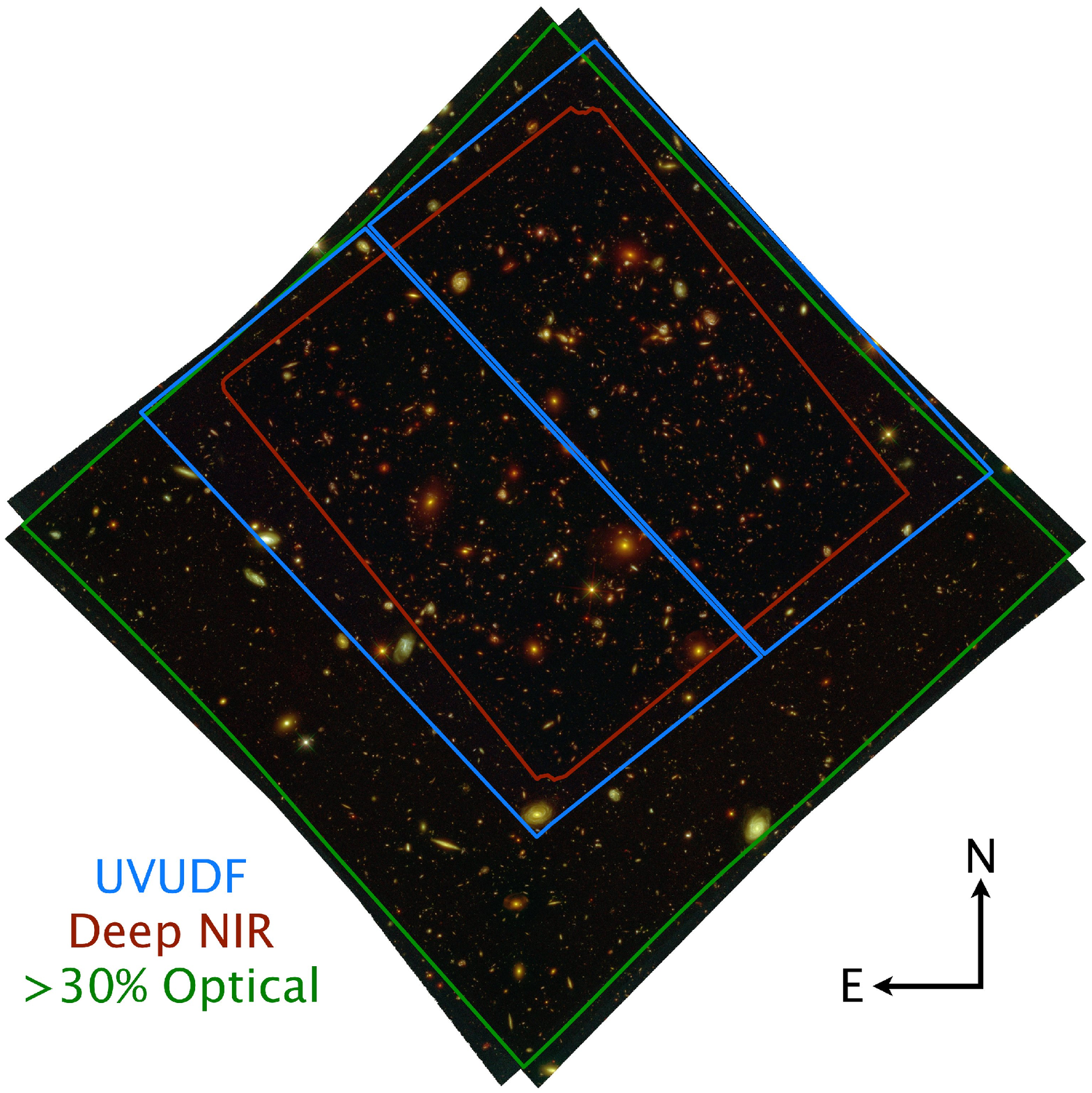}
}
\caption{ \label{fig:coverage} 
UDF coverage maps for the WFC3-UVIS, ACS-WFC, and WFC3-IR data overlaid on a color image created from all eleven bandpasses. 
The entire field is imaged in the original four ACS optical filters \citep{Beckwith:2006} and the three WFC3-IR filters from CANDELS \citep{Grogin:2011,Koekemoer:2011}. 
The UV coverage for the three WFC-UVIS filters \cite{Teplitz:2013} is outlined by the blue rectangles, 
and the deep IR region for the four WFC3-IR filters is outlined by the red rectangle \citep{Oesch:2010a,Oesch:2010b,Bouwens:2011b, Koekemoer:2013, Ellis:2013}.
The green rectangle outlines the area covered by more than 30\% exposure time in the optical UDF.
North points upwards in the image.
}
\end{figure}

This study makes use of eleven photometric bandpasses covering the UDF ($\alpha(J2000) = 03^{h}32^{m} 39^{s}$, $\delta(J2000) = -27^{\circ}$47$\tt'$29.$\tt''$$1$) at high spatial resolution, spanning wavelengths from the NUV to the NIR. 
The blue box in Figure \ref{fig:coverage} outlines the NUV coverage of the UDF (UVUDF), which is comprised of three WFC3-UVIS filters:  F225W, F275W, and F336W, as described by \citep{Teplitz:2013}. 
These data have been modified from those discussed in \citet{Teplitz:2013} to include improved calibrations, photometry, and astrometry as described below. 
The optical data defines the field region in the Figure, and is covered by the four original ACS optical filters: F435W, F606W, F775W, and F850LP \citep{Beckwith:2006}.
The red box in the Figure outlines the deep NIR coverage, which includes four WFC3-IR filters: F105W, F125W, F140W, and F160W obtained in the UDF09 and 
UDF12 programs \citep{Oesch:2010a,Oesch:2010b,Bouwens:2011b, Koekemoer:2013, Ellis:2013}. 
The entire field is also covered by three of the four WFC3-IR filters (F105W, F125W, and F160W) in the CANDELS GOODS-S observations \citep{Grogin:2011, Koekemoer:2011}.
The full UDF field, NUV coverage, and deep NIR coverage cover areas of 12.8 arcmin$^2$, 7.3 arcmin$^2$, and 4.6 arcmin$^2$ respectively. The full UDF coverage area shrinks to 11.4 arcmin$^2$
when considering only the area covered by more than 30\% of the optical exposure time, as shown by the green box in the Figure.

\begin{figure}[t!]
\center{
\includegraphics[scale=0.5, viewport=15 5 500 360,clip]{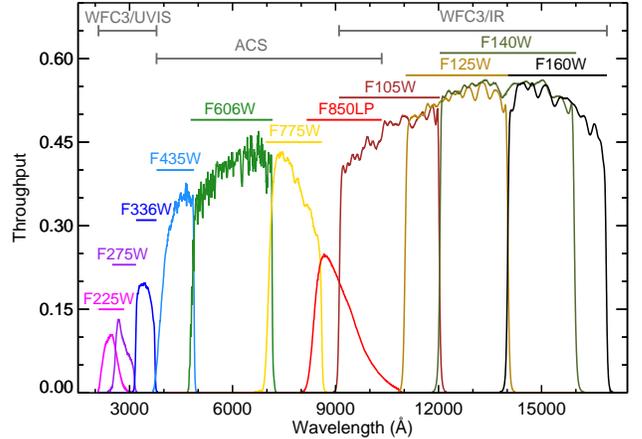}
}
\caption{ \label{fig:filters} Total system throughput of the WFC3-UVIS, ACS-WFC, and WFC3-IR filters used
  in the UDF. Each filter is plotted as a separate color, with lines of the same color denoting the wavelength range sampled. 
  The gray bars at the top show the wavelength range used for each of the three instruments. Note that these
  throughputs include the quantum efficiency of the CCD. }
\end{figure}

We do not use the NIR NICMOS J and H data \citep{Thompson:2005} or the WFPC2 F300W observations \citep{Voyer:2009} in our analysis because the same wavelengths are covered by other filters presented here with superior depth, resolution, and coverage area. Also, in general we do not include lower resolution observations of the UDF, such as ground-based imaging \citep[e.g.,][]{Rafelski:2009} or low-resolution space-based imaging such as Spitzer IRAC \citep[e.g.,][]{Ashby:2013}, to avoid source confusion, correlated photometry, and the added systematic uncertainties associated with more complex photometric techniques (see Section \ref{aptpsf}). Lastly, the SBC far-UV data \citep{Siana:2007} are also not used here as they contain very few sources and thus would not significantly affect the results. The filter coverage and throughput for all eleven bandpasses is shown in Figure \ref{fig:filters}. The large number of filters over such a large wavelength range enables the simultaneous measurement of the Lyman Break and the Balmer/4000\AA ~break at redshifts $0.8\lesssim z\lesssim 3.4$, significantly improving the photometric redshifts of these galaxies (see section \ref{photoz}). 

\subsection{General UDF Imaging Information}
\label{udfsum}

The data represent the deepest high-resolution panchromatic dataset available, although the properties of each bandpass varies by instrument and filter. Table \ref{tab:obs} provides information about each bandpass, including the effective wavelength, zero point, Galactic extinction, exposure time, depth, areal coverage, and point spread function (PSF) full width at half maximum (FWHM). The Galactic extinction is derived from the extinction coefficients in \citet{Postman:2012},  using E(B$-$V) = 0.00782 based on the \citet{Schlegel:1998} IR dust-emission maps. The PSF FWHM is measured by fitting a symmetrical Gaussian to the PSFs as described in Section \ref{psf}. Although the PSFs are not Gaussian, these provide estimates of the resolution of the images for reference purposes.

The depth measurements in Table \ref{tab:obs} are obtained by measuring the 5$\sigma$ sky noise of each image in the same fashion as \citet{Teplitz:2013}; the sky rms is measured in 1000 semi-random empty $51\times51$ pixel boxes, multiplied by the correlation ratio for each mosaic from \citet{Fruchter:2002}, and normalized to an aperture of 0.2\arcsec radius. All the data are drizzled to a pixel scale of 0.03 arcsec pixel$^{-1}$, and therefore the correlated noise is highest in the NIR (largest pixels; 0.128 arcsec pixel$^{-1}$), followed by the optical (0.05 arcsec pixel$^{-1}$) and then the NUV (0.0396 arcsec pixel$^{-1}$). The aperture choice of 0.2\arcsec radius is optimized for relatively compact galaxies in the optical and NUV, and slightly overestimates the depths of the NIR data.

\begin{deluxetable*}{llcccccccccccccccccccccccccccc}
\tablecaption{UDF Imaging Summary \label{tab:obs}}
\tablehead{
\colhead{Instrument/} &
\colhead{Filter} &
\colhead{Effective\tablenotemark{a}} &
\colhead{Zero} &
\colhead{Galactic } &
\colhead{Number} &
\colhead{Exposure} &
\colhead{$5\sigma$\tablenotemark{b}} &
\colhead{Area} &
\colhead{PSF} \\
\colhead{Camera} &
\colhead{} &
\colhead{Wavelength} &
\colhead{Point} &
\colhead{Extinction} &
\colhead{of orbits} &
\colhead{Time} &
\colhead{Depth} &
\colhead{ } &
\colhead{FWHM} \\
\colhead{} &
\colhead{} &
\colhead{(\AA)} &
\colhead{(ABMAG)} &
\colhead{(ABMAG)} &
\colhead{} &
\colhead{(s)} &
\colhead{(ABMAG)} &
\colhead{(arcmin$^2$)} &
\colhead{(arcsec)}
}
\startdata

WFC3/UVIS & F225W  & 2359 & 24.0403 & 0.058   & 16 &44,072 & 27.8 & 7.3 &0.11 \\
WFC3/UVIS & F275W  & 2704 & 24.1305 & 0.048   & 16 &41,978 & 27.8  & 7.3 &0.10 \\
WFC3/UVIS & F336W  & 3355 & 24.6682 & 0.040   & 14 &37,646 & 28.3  & 7.3 &0.09 \\
ACS/WFC & F435W  & 4317 & 25.673  & 0.033 & 56 &134,880& 29.2  & 12.8\tablenotemark{e} &0.10 \\
ACS/WFC & F606W  & 5918 & 26.486  & 0.023 & 56 &135,320& 29.6  & 12.8\tablenotemark{e} &0.10 \\
ACS/WFC & F775W  & 7693 & 25.654  & 0.016 & 144 &347,110& 29.5  & 12.8\tablenotemark{e} &0.09 \\
ACS/WFC & F850LP & 9055 & 24.862  & 0.012 & 144 &346,620& 28.9  & 12.8\tablenotemark{e} &0.10 \\
WFC3/IR\tablenotemark{c} & F105W  & 10550 & 26.2687 & 0.0079  & 100 &265,459& 30.1  & 4.6 &0.19 \\
\tablenotemark{d} &   &  &  &  & 4 &8,100 & 28.7  & 8.2\tablenotemark{f} & \\
WFC3/IR\tablenotemark{c} & F125W  & 12486 & 26.2303 & 0.0069  & 39 &226,586& 29.7  & 4.6 &0.19 \\
 \tablenotemark{d}&  &  &  & & 5 &9,100 & 28.6  & 8.2\tablenotemark{f} & \\
WFC3/IR\tablenotemark{c} & F140W  & 13923 & 26.4524 & 0.0048  & 30 &82,676 & 29.8  & 4.6 &0.20 \\
WFC3/IR\tablenotemark{c} & F160W  & 15370 & 25.9463 & 0.0037  & 84 &352,674& 29.9  & 4.6 &0.20 \\
\tablenotemark{d} &  &  &  & & 5 &9,200 & 28.2  & 8.2\tablenotemark{f} &

\enddata
\tablecomments{UDF image mosaic information, including filters, zero points, extinction, exposure times, sensitivities, and area coverage for each of the eleven bandpasses.}
\tablenotetext{a}{Effective wavelength as calculated in \citet{Tokunaga:2005}, also known as the `pivot wavelength'.}
\tablenotetext{b}{Limiting $5\sigma$ depth base on sky noise in empty regions of the detector in an aperture of 0.2\arcsec radius.}
\tablenotetext{c}{Deep IR data from UDF09 and UDF12 surveys \citep{Oesch:2010a,Oesch:2010b,Bouwens:2011b, Koekemoer:2013, Ellis:2013}}
\tablenotetext{d}{Shallow IR data from CANDELS \citep{Grogin:2011,Koekemoer:2011}}
\tablenotetext{e}{Source catalog is trimmed to central 11.4 arcmin$^2$, see Section \ref{sources}).}
\tablenotetext{f}{Source catalog trimming results in this covering 6.8 arcmin$^2$, see Section \ref{sources}). }
\end{deluxetable*}

\subsection{NUV Data}
\label{newnuv}

The UVUDF observations were obtained in 2012 using WFC3/UVIS in three filters, two different observing modes, and three orientations as described in \citet{Teplitz:2013}. The first half of the data (referred to as Epochs 1 and 2) was obtained in $2\times2$ binned mode, while the second half (Epoch 3) was obtained in the unbinned mode, with the addition of ``post-flash'' to add internal background light to the image to mitigate charge transfer efficiency (CTE) degradation (see Section \ref{cte}). While these data are described in detail in \citet{Teplitz:2013}, here we present improvements to the processing of the NUV data that significantly improve the final image mosaics. This includes a CTE correction, new astrometric alignment, and new dark calibrations. In addition to the improved darks and other improvements, we applied a background subtraction to each science exposure as described in \citet{Teplitz:2013}, to ensure that no leftover gradient is present in the final image mosaics. These data have been released on the Barbara A. Mikulski Archive for Space Telescopes (MAST) website at \url{http://archive.stsci.edu/prepds/uvudf/}. They cover $\sim60\%$ of the UDF FOV. 

\subsection{Charge Transfer Efficiency}
\label{cte}

The CTE of the WFC3/UVIS detector has significantly degraded over time, as radiation causes permanent damage of the charge transfer device (CCD) lattices  \citep{MacKenty:2012}. This damage degrades the ability of electrons to transfer from one pixel to another, temporarily trapping electrons during the readout. When uncorrected, the electrons are smeared out in the readout direction, appearing as trails in the images. This affects the photometry and measured morphology of the objects in the images \citep{Rhodes:2010, Massey:2010a}. CTE degradation is most severe for low-background imaging of faint sources, such as NUV imaging and calibration dark frames, where faint sources or hot pixels can be lost completely \citep{Anderson:2012c}. The effects of CTE degradation on the UVUDF data products without any corrections are described in detail in \citet{Teplitz:2013}, who find that these issues are somewhat mitigated  by the use of ``post-flash'', thereby reducing the smearing effect, and avoiding the loss of faint sources and hot pixels.

For this paper, only the post-flashed data are used, which significantly reduces the effects of CTE degradation. The post-flashed data are from ``Epoch 3'' and include $\sim$13 electrons per pixel background, \citep[see][]{Teplitz:2013}. In addition, we apply a pixel-based CTE correction\footnote{\url{http://www.stsci.edu/hst/wfc3/tools/cte\_tools}} to the raw data based on empirical modeling of hot pixels in dark exposures \citep{Anderson:2010, Massey:2010a}. This correction not only corrects the photometry, but also restores the morphology of sources. Currently, this software is only available for unbinned data, and thus the Epoch 1 and 2 data described in \citet{Teplitz:2013} are not included in our analysis.  While this reduces the overall depth of the NUV data, the CTE correction is critical to obtaining accurate photometric and morphological measurements.  CTE corrections are unnecessary for the optical and NIR data, as the ACS/WFC optical data were obtained early in the lifetime of the ACS/WFC and the HgCdTe NIR detector onboard WFC3/IR is not affected by CTE degradation.


\subsection{WFC3/UVIS Dark Calibrations}

Dark calibrations are especially important for NUV data because the dark current level in each exposure is high relative to the low sky background. In addition, regular calibration dark data can be used to identify hot pixels, which vary significantly over time. \citet{Teplitz:2013} show that the darks currently provided by STScI are insufficient for data with low background levels after the CTE degradation of WFC3/UVIS. In this paper we improve the dark calibrations even further, as discussed below. 

While the STScI superdarks were mostly sufficient for early data obtained soon after the installation of WFC3, subsequent changes in the characteristics of the detector (such as CTE degradation) increasingly affected the science data. There are three major areas that the STScI processed superdarks are insufficient for use in the UDF program. First, when the darks produced by STScI are used to calibrate the data, more than half of the hot pixels are missed. Second, an observed background gradient is left unremoved, and finally, a blotchy pattern is left in the images. We discuss these three effects in Appendix A. They are present even when the post-flashed raw dark data are corrected with the pixel-based CTE correction.

In Appendix A we present a new dark processing methodology which improves on the previous work by STScI and \citet{Teplitz:2013}. This new methodology is being used by a large number of other HST programs (e.g., PI: Faber \& Ferguson 12444, 12445, PI:  Siana, 12266, 12201, 13389, PI: Guo 13309, PI: Malkan, 12283, 12568, 12902, 13352, 13517), and we describe the methodology in detail here. A similar strategy is being developed at STScI, and will be implemented in their calibration pipeline in the future. 

\subsection{Astrometric Alignment}

Here we discuss several sources of astrometric uncertainties in the original data, as well as our approaches to mitigating these and aligning all the images to a common reference grid. The observations were all obtained in a non-integer pixel-offset dither pattern, aimed at ensuring that the point spread function (PSF) was adequately sampled in the final mosaics. Due to the geometrical distortion of the detector, shifts that are too large correspond to a substantially different number of pixels along the edge than at the center where the shifts are defined, which would cause the pixel subsampling phase to change across the detector. Therefore the shifts are kept small enough to retain the intended sub-pixel subsampling across the detector. Each small angle maneuver introduces a slight uncertainty in positioning (of the order of about 1 to 2 milliarcseconds). In addition, an optical offset is introduced when a different filter is inserted into the optical path. Moreover, during each orbit the spacecraft undergoes thermal expansion and contraction (``breathing'') due to changes in solar illumination, which lead to changes in the optical path length to the detectors, hence resulting in slight scale changes from one exposure to the next. Finally, guide star reacquisition uncertainties can lead to errors in position as well as small rotation uncertainties, while a full acquisition of a new guide star has astrometric uncertainties of $\sim\,$0$\farcs$3$\,-\,$0$\farcs$5 (reflecting the absolute astrometric uncertainties in the Guide Star Catalog 2).

We make use of the source positions measured in the F435W ACS mosaics of the UDF \citep{Beckwith:2006} as our absolute astrometric reference frame. In this way, we minimize astrometric scatter resulting from color terms across sources. Initially we aligned the F336W to this frame, which then enabled us to bootstrap the F275W data, and subsequently the F225W data. The astrometric accuracy also depends on the accuracy of distortion calibration for WFC3; for this program, we found significantly improved results when using an updated distortion model made publicly available by the WFC3 team which was implemented in the archive in 2013.
The initial alignment was accomplished with {\tt drizzlepac/tweakreg} using catalog matching, which provides measurements of rotations as well as removing the bulk of the shifts that are present in the data. Further details of this technique are presented in \citet{Koekemoer:2011}, updated with new distortion models that are available within {\tt drizzlepac/astrodrizzle}. Once all the rotations and the majority of the shift information is solved for, the images were then passed through astrometric refinement using cross-correlation, which provides additional improvements, resulting in overall astrometric accuracies of $\sim$2 milliarcseconds in the mean shift positions of all the exposures relative to one another, which is the best possible level that is achievable given the sparse number of sources and their faint flux at these wavelengths, particularly in F225W.

\subsection{Extended UDF Near-IR Mosaics}

The UDF12 program \citep{Koekemoer:2013, Ellis:2013} originally released mosaics of the NIR UDF at the 60 mas scale, covering the extremely deep pointing of the NIR UDF from both the UDF12 and the UDF09 programs \citep{Oesch:2010a,Oesch:2010b,Bouwens:2011b}. The initial release of these mosaics do not cover the whole field of view of the original UDF or the new NUV observations (see Figure \ref{fig:coverage}). In addition, given the higher resolution and smaller plate scale of the NUV data, the NUV data mosaics are released at the 30 mas scale to MAST, enabling  improved morphological measurements in the UDF \citep[e.g.,][]{Bond:2014}. 

To facilitate photometry at the same plate scale as the NUV and optical data, we make use of new 30 mas scale NIR mosaics that include data from the surrounding CANDELS survey \citep{Grogin:2011, Koekemoer:2011} as well as the full WFC3/IR dataset on the UDF footprint \citep{Koekemoer:2013, Ellis:2013}. In this way, the entire original optical UDF is covered by NIR data, although at a shallower depth in the regions outside of the UDF09/UDF12 NIR pointing. The depths and covered areas are listed in Table \ref{tab:obs}. The area of the shallower NIR data covers almost twice the area of the deep NIR pointing, a significant contribution to the UDF dataset. Moreover, while the data are shallower than the deep pointing, their depth is similar to our NUV data depths, and significantly improve the photometric redshifts we describe below. 

\subsection{Point Spread Functions}
\label{psf}

In order to measure PSF-corrected aperture-matched photometry as described below, PSFs of the different bandpasses are required. PSFs for each HST camera are created in slightly different fashions, due to varying constraints of the data. 

The WFC3/NIR data PSFs are created in a very similar manner to the PSFs created by CANDELS \citep{vanderWel:2012}. Specifically, the NIR PSFs are hybrid PSFs, with the central pixels and the outer wings based on a model PSF, and the middle portion consisting of a stack of stars in the UDF. This hybrid approach is necessary since the large wings of the NIR PSFs contain approximately $\sim10$\% of the flux beyond 1\arcsec (WFC3 instrument handbook), which is difficult to include solely by stacking the limited number of stars in the UDF. In addition, the NIR models were found to not accurately reproduce the PSF at intermediate radii compared to stars in CANDELS, thereby justifying a hybrid PSF over using a pure model PSF \citep{vanderWel:2012}. Note that the CANDELS PSFs can not be used directly, because we used a drizzle scale of 30 mas rather than 60 mas.

The NIR PSF model is created with the {\tt TinyTim} package \citep{Krist:1995} for the center of the WFC3 detector, sub-sampled to align the planted PSFs, re-sampled to the WFC pixel scale, distortion corrected, and then combined with the same dither pattern and drizzle parameters as was used to produce the imaging mosaics. The resultant models are similar to those from \citet{vanderWel:2012}, except at a 30 mas scale. The NIR stack of stars is created from 7 unsaturated stars in the UDF deep NIR region (Figure \ref{fig:coverage}) selected from the published catalog of stars in the UDF \citep{Pirzkal:2005} such that they are not contaminated in the wings by nearby galaxies. In addition, one star is excluded as it has a slightly extended profile and it is likely an AGN. The stars are registered to their subpixel centers, normalized by the peak value, and then coadded via a median. The final hybrid PSF is a combination of the two, composed of the PSF model from a radius of 0-5 pixels (0-0.15\arcsec), of the stack of stars from 5 pixels to 35 pixels (0.15-1.05\arcsec), and of the PSF model from 35-166 pixels (1.05-5.0\arcsec).

The WFC/ACS optical PSFs are median stacks of stars, constructed from 15 stars for the F435W and F606W bands, and 8 stars for F775W and F850LP bands selected from the published catalog of stars in the UDF \citep{Pirzkal:2005}. The stars are registered to their subpixel centers, normalized by the peak value, and then coadded via a median. The optical bandpasses have a smaller PSF, with less flux in the wings and a narrower central peak, so we did not need to extend their PSFs to larger radii via a PSF model. 

The WFC3/UVIS PSFs are created with a hybrid PSF in a fashion similar to the WFC3/IR PSFs. The PSF FWHM of the NUV data is similar to the optical data, with $<5$\% of the PSF flux beyond 1\arcsec (WFC3 instrument handbook). However, due to the low number of stars in the NUV (2-3), the wings of the PSFs can not be recreated from a stack of stars. In fact, the best PSF in the NUV is generated by a single bright star. Coadding this higher SN star with the 1 or 2 low SN stars in the data does not improve the SN of the final PSF star in the NUV. Therefore, the best PSF is generated by combining a single high SN star with a PSF model. 

The hybrid NUV PSFs are created in a similar fashion as the NIR PSFs described above, with a few differences. The model PSF is only used in the wings, as the central part of the PSF is sufficiently constrained by the single star due to the higher resolution and smaller plate scale. At the same time, the model is incorporated at a smaller radius due to lower SN in the wings. In order to prevent discontinuities in the resultant PSF, the PSF model and the star are added together, weighted by a Gaussian with a full width of 15 pixels (0.45\arcsec).

The FWHM values for all the PSFs are listed in Table \ref{tab:obs}. For this measurement, a symmetric Gaussian is fit to the final PSFs, although that is not a good fit to these non-Gaussian PSFs. However, the Gaussian fit only provides an estimate of the size of the FWHM for the table. The NUV PSF FWHM's measured here are smaller than those measured by \citet{Teplitz:2013} in the same data, which is expected because the data in \citet{Teplitz:2013} were not yet corrected for CTE degradation, which artificially made the PSFs more extended. The resulting PSF FWHM's are similar to those measured by \citet{Windhorst:2011}, although slightly larger. While PSFs are created for all the images, only the F775W and the four NIR PSFs are used in this paper, as described in Section \ref{aptpsf}.

\section{Photometry and Catalog Creation}
\label{photometry}
\subsection{Aperture-Matched PSF Corrected Photometry}
\label{aptpsf}
Aperture-matched PSF-corrected photometry is essential for robust photometry of galaxies using image data with varying PSFs \citep[e.g.,][]{Benitez:1999, Vanzella:2001, Coe:2006, Laidler:2007, deSantis:2007, Finkelstein:2012a}. As the PSF FWHM increases, the accuracy of galaxy photometry is affected. If sufficiently large apertures are used, then the photometry of bright galaxies are not significantly affected by this effect. However, for faint galaxies, a significant fraction of the flux is under the level of the noise, resulting in large differences in photometry regardless of the aperture size. Multiple methods are possible to correct the photometry for the PSF variation, with the three most prominent ones discussed here.

First, if the PSF FWHM's are very different (by a factor of a few or more), then a template-fitting method is typically used to match a convolved galaxy `model' to the measured flux \citep{Laidler:2007, deSantis:2007}. Since our analysis is limited to data with similar PSF FWHM's (within a factor of two), a simpler approach can be used. 
A second option is to convolve all the images to the PSF of the largest PSF image \citep[e.g.,][]{Grazian:2006,Finkelstein:2012a}, and measure photometry on these PSF matched images. The downside of this approach is that all measurements are then performed on convolved data rather than the original data, at the worst resolution of the images, and any systematic uncertainties in the PSFs will affect the photometry in all bands. This is particularly a problem for compact faint galaxies and low SN observations, for which the SN is significantly degraded once they are convolved by a larger PSF because the convolution includes pixels that are mostly noise.

The third option, and the one used here, is to measure the photometry in the original higher-resolution data, and then apply a PSF correction on the NIR measurements, which have larger PSF FWHMs, as done in \citet{Coe:2006}. Since the PSFs of the NUV and optical data are similar, PSF corrections between these bandpasses could introduce systematic uncertainties based on the quality of the PSFs, which could be as large as the corrections themselves. More importantly, this method enables the use of a separate detection image for the NUV data as described in Section \ref{bband}, which improves the SN in the NUV data by using smaller apertures, and afterwards applying an aperture correction. Since the emphasis of this paper is the inclusion of the NUV data, the third method is applied in this paper. 

\begin{deluxetable*}{lrrrrrr}
\tablecaption{Source Extractor Parameters for Different Runs \label{tab:apertures}}
\tablehead{
\colhead{Parameter} &
\colhead{Deep} &
\colhead{Shallow} &
\colhead{Deep Deblend\tablenotemark{a}} &
\colhead{Shallow Deblend\tablenotemark{a}} &
\colhead{NUV} &
\colhead{NUV Deblend\tablenotemark{a}} }
\startdata
{\tt detect\_thresh} & 1.1 & 3.5 & 1.1 & 3.5 & 1.0 & 1.0 \\
{\tt analysis\_thresh} & 1.1 & 3.5 & 1.1 & 3.5 & 1.0 & 1.0 \\
{\tt deblend\_nthresh} & 32 & 32 & 8 & 8 & 32 & 8 \\
{\tt deblend\_mincont} & 0.01 & 0.01 & 0.3 & 0.3 & 0.01 & 0.3 \\
{\tt detect\_minarea} & 9 & 9 & 9 & 9 & 9 & 9 \\
{\tt back\_size} & 128 & 128 & 128 & 128 & 128 & 128 \\
{\tt back\_filtersize} & 5 & 5 & 5 & 5 & 5 & 5 \\
{\tt back\_photo\_thick} & 26 & 26 & 26 & 26 & 26 & 26 
\enddata
\tablecomments{Table of {\tt SExtractor} parameters used for the different runs to create the catalog.
}
\tablenotetext{a}{Deblend represents the {\tt SExtractor} runs with a low deblending threshold. }
\end{deluxetable*}

\subsection{Overview of Methodology}
We first summarize the general methodology, and then provide details in the following sub-sections. 
We use {\tt ColorPro} to measure photometry in the images \citep{Coe:2006}, which is a wrapper for running Source Extractor ({hereafter \tt SExtractor}) \citep[v2.5.0;][]{Bertin:1996} on multiple images. Each run of {\tt ColorPro} is analogous to running {\tt SExtractor} in dual-image mode eleven times, once for each filter, and also applies aperture and PSF corrections (see Section \ref{psfcorr}). All photometry is therefore determined by running {\tt SExtractor} in dual-image mode on both a detection image and each individual image. The detection image is created from multiple images obtained with different filters, to maximize its depth and the robustness of the aperture sizes (see Section \ref{detect_image}). 

A single detection threshold that optimizes the apertures of most of the objects in the images results in apertures that are poorly defined for both the bright sources, and fainter sources near them. The solution is to determine apertures with two different detection thresholds. In addition, for each detection threshold, we also measure the photometry using two different deblending thresholds, as no single threshold sufficiently deblends some objects, without artificially splitting up single objects. This results in 4 separate measurements for each galaxy, with differing object definitions and aperture sizes, which are combined into a single photometric catalog and segmentation map (see Section \ref{sources}).

To optimize the SN in the NUV data, smaller apertures are needed, and are determined by running {\tt SExtractor} in dual image mode with the F435W image, instead of the general detection image, for the three NUV images. Two separate deblending thresholds are required, and the resulting NUV photometry is merged with the full 11-band catalog, including an aperture correction for the different aperture sizes (see Section \ref{bband}).

\subsection{Aperture and PSF corrections}
\label{psfcorr}

The PSF correction is determined by degrading the F775W image to the PSF of each of the 4 NIR images using the IRAF task {\tt psfmatch}. The F775W image is used as the high-resolution image, as it is the reddest high-resolution image with a well-behaved PSF \citep[for a discussion of F850LP, see][]{Coe:2006}. For each image, a PSF Kernel is created, which if convolved by the F775W image, matches the PSF of the corresponding NIR image. The {\tt psfmatch}  {\tt threshold} parameter, which sets the low frequency cutoff in the Kernel creation, is set to 0.14 to minimize fringing in the resulting convolved image. 

{\tt ColorPro} measures the colors of galaxies based on their isophotal fluxes, which have been shown to produce robust colors, outperforming circular or large Kron apertures \citep{Benitez:2004}. It then applies an aperture correction from the ratio of the isophotal flux and the total flux measured via the {\tt SExtractor} {\tt mag\_auto} Kron \citep{Kron:1980} aperture flux. In addition, for the small number of sources in which an aperture and PSF correction cannot be calculated, ColorPro calculates it based on other sources with similar sizes (and thereby also magnitude).

\subsection{Multi-band Detection Image}
\label{detect_image}

The detection image determines the object definitions and aperture sizes, and thus a very sensitive image is desired. It is also important for the apertures to be sufficiently large to contain the majority of the light in the worst resolution image, yet still be sensitive to smaller faint objects that may not be detected in the images with larger PSF FWHMs. A well-matched aperture minimizes the effects of any internal color gradients of the galaxies. Therefore, we make the detection image by averaging the 4 optical images with the 4 NIR images together, weighted by the inverse variance of each image on a pixel-by-pixel basis. In this fashion, the varied depth of the NIR image is taken into account in the region covered by the shallow NIR data, and the resultant RMS image is outputted for use with {\tt SExtractor}. This detection image optimizes source detection at both resolutions as we do not lose the faint compact sources present in the optical data but not in the NIR data. Although averaging two different resolutions together creates slight 'donut' images, this is sufficiently minor that it is not visible in the image, and {\tt SExtractor} is not sensitive to this when defining the segmentation map as the detection image is convolved by a Gaussian in the {\tt SExtractor} run. The alternative method of first smoothing all the data to the resolution of the largest PSF would reduce the SN of faint compact sources, losing these potentially interesting sources. This detection image methodology is similar to that used by the CLASH team \citep{Postman:2012,Coe:2013}.

\subsection{Source Aperture Definitions}
\label{sources}

While there are many options available to optimize {\tt SExtractor} to obtain the best object definitions possible, no single set of detection thresholds and deblending parameters perfectly detects bright, faint, large, and small galaxies simultaneously. Either a lot of galaxies are blended together, or the larger galaxies are split into multiple segments. This is an intrinsic limitation with the current automation methods, although for the vast majority of sources, source definitions are adequate. To solve these issues for the problematic sources, {\tt SExtractor} is run four times on each image, with each run optimized differently, and then merged together into a final catalog as described below. Two are optimized via the detection threshold, and two include different deblending thresholds. This is similar to the 'hot' and 'cold' mode method by \citet{Barden:2012}, with differences in the methodology being due to independent development. The different parameters are also summarized in Table \ref{tab:apertures}.

The first iteration is a `deep' {\tt SExtractor} run, with the {\tt detect\_thresh} and {\tt analysis\_thresh} parameters set to 1.1$\sigma$, which determines the minimum deviation above the background RMS (including correlation corrections for the drizzle). A minimum of 9 contiguous pixels above this threshold are required to define a source as set by {\tt detect\_minarea}. The  background sky for detection is computed by setting {\tt back\_size} to 128 and {\tt back\_filtersize} to 5, and the local sky for photometry is determined with a background annulus of 26 pixels ({\tt back\_photo\_thick}). These parameters yield reasonable performance in detecting faint sources while minimizing spurious detections. 

With the above threshold, however, the source definitions near bright targets are poorly defined for both the bright sources, and fainter sources near them. The solution is to run {\tt SExtractor} with a higher detection threshold, with {\tt detect\_thresh} set to 3.5$\sigma$. This second `shallow' run detects significantly fewer sources, although the bright source detections are well defined. The two catalogs are then combined by replacing the bright sources and their neighbors in the first run with the shallow run measurements. Sources brighter than 22nd magnitude are replaced, which includes 55 bright sources and 1014 fainter neighbors, which fall within 250 pixels of the brighter sources. Faint sources which are not detected in the shallow run within 250 pixels are also included, to avoid losing any sources.

Both of these {\tt SExtractor} runs have difficulty deblending sources properly; either the deblending thresholds are set too low, and multiple objects are blended together, or the thresholds are set too high, and single galaxies are deblended into multiple sources based on knots or spiral structure in the galaxies. The solution implemented here is to run {\tt SExtractor} twice; first with`normal' deblending thresholds optimized for the vast majority of galaxies, with {\tt deblend\_nthresh} set to 32 and {\tt deblend\_mincont} to 0.01, and second with `low' deblending thresholds optimized for larger galaxies, with {\tt deblend\_nthresh} set to 8 and {\tt deblend\_mincont} to 0.3. The threshold dependent source definitions determined in the first two runs are manually checked for galaxies that should not be deblended, and replaced by these 3rd and 4th runs. This only affects 90 galaxies in the final catalog, although a larger number of galaxies are deblended and should not be. This is due to the conservative approach used, which only combined galaxies that are obviously incorrectly deblended to avoid combining multiple individual galaxies together. A small number galaxies can not be combined together with this technique, even when clearly they should be, as other nearby galaxies would be incorrectly deblended in those cases. Hence, some galaxy blending and deblending issues will be in the final catalog. The resulting catalogs are all merged into a single catalog, with a single segmentation map. This is then fed back into {\tt ColorPro} to create a final catalog ready for calculating photometric redshifts.

\subsection{F435W Apertures For NUV Photometry}
\label{bband}

One of the primary goals of this paper is to include the new NUV observations of the UDF \citep{Teplitz:2013} in the galaxy photometry. While the galaxies in the data have similar rest-frame optical sizes \citep{Bond:2014}, the images are somewhat shallower than the rest of the UDF data (see Table \ref{tab:obs}), and the galaxies in the NUV are fainter due to their spectral energy distribution (SED). While the apertures determined in Section \ref{sources} could work for the NUV data, the SN is not optimized when using very large apertures on small faint objects.   In addition, the use of smaller apertures helps avoid remaining calibration issues observed as a blotchy pattern in the dark calibrations which results in an increased flux and uncertainty over very large areas (see Section \ref{blotchyscience}). Therefore, smaller apertures are used for the NUV data determined from the F435W mosaic of the UDF, enabling more robust and higher SN measurements.

{\tt SExtractor} is run in dual image mode with the F435W mosaic as the detection image for all three NUV mosaics and the F435W image. Since sources in the F435W image are fainter than in the redder bands, a single threshold selected such that no sources are detected in the negative image is sufficient, with {\tt detect\_thresh}$=$1.0$\sigma$ and {\tt detect\_minarea}$=$9. By requiring no sources are detected in the negative image with the same {\tt SExtractor} parameters, we ensure that the parameters are optimized since the negative image has the same noise properties as the normal image, but with no real sources. In this way there will only be almost no false sources detected by {\tt SExtractor} when run on the original images, except by potential image artifacts \citep{Rafelski:2009}.

Galaxies in the F435W band are clumpier and have a higher level of apparent structure in the images \citep{Guo:2015}, making two runs with different deblending thresholds more important than in the multi-band detection image. The same two deblending thresholds are used as in Section \ref{sources}, with a normal and low deblending threshold. Sources detected in the low deblending threshold case are cross-checked with the merged multi-band detection segmentation map from Section \ref{sources}. If the segmentation map contains more than a single segment in the area covered by the smaller F435W segment area, then the normal deblending threshold is used as otherwise multiple galaxies are likely to be blended together. Otherwise, the low deblending threshold is used, to maximize the combination of clumps and other structure in galaxies into single galaxy detections. In this way, the F435W source definitions are similar to the source definitions of the multi-band detection definitions.

The new F435W detected sources are then matched to the multi-band detected sources using the merged segmentation maps described above. For each source in the multi-band catalog, all the pixels in the aperture defined by the segmentation map are checked in the F435W segmentation map. If only a single source is found within that aperture, then the source is matched. If two sources fall within the aperture and the second object covers 40\% or more of the area of the source, then the two sources are added together and errors propagated. Otherwise, only the main source is matched to the multi-band catalog. If no source is detected in the F435W band within the detected area, it is marked as undetected in the catalog. Since the NUV data are significantly shallower than the F435W band, all sources not detected in F435W are assumed not detected in the NUV bands, which will almost always be the case. This methodology works well because if there are multiple sources in the aperture, then they were detected with the low deblending threshold parameters, and most galaxies are either one or two sources before being matched. In addition, if a source is close to the edge or the chip gap of the NUV data, it is marked as not covered. 

Since the F435W detected apertures are smaller than the multi-band apertures, an aperture correction is added to place it on the same magnitude scale as the larger aperture longer wavelength data. The aperture correction is determined from the higher SN F435W image, consisting of the difference in the large aperture and small aperture magnitudes. As with all aperture corrections, we assume the color gradient from F435W to the NUV is small. 

\begin{deluxetable*}{lccrrr}
\tablecaption{Spectroscopic Redshift Compilation \label{tab:specz_comp}}
\tablehead{
\colhead{Survey\tablenotemark{a}} &
\colhead{Telescope\tablenotemark{b}} &
\colhead{Instrument\tablenotemark{c}} &
\colhead{Number} &
\colhead{Quality} &
\colhead{Reference}}
\startdata
VVDS &  VLT  & VIMOS  & 8 & 95\% & \citet{LeFevre:2004} \\
Szokoly & VLT & FORS1/FORS2 & 4 & `reliable' (2 or 2+) & \citet{Szokoly:2004} \\
K20 & VLT & FORS1/FORS2 & 27 & `secure' (1) & \citet{Mignoli:2005} \\
GRAPES & HST & ACS & 6 & UV absorption lines & \citet{Daddi:2005ek} \\
Vanzella GOODS  & VLT & FORS2/VIMOS & 39 & A or B multiple lines & \citet{Vanzella:2005, Vanzella:2006, Vanzella:2008, Vanzella:2009} \\
Popesso GOODS & VLT & VIMOS & 19 & A or B multiple lines & \citet{Popesso:2009} \\
Balestra GOODS & VLT & VIMOS & 18 & A or B multiple lines & \citet{Balestra:2010} \\
GMASS & VLT & VIMOS & 29 & `good' (1) & \citet{Kurk:2013} \\
3D-HST & HST & WFC3 & 28 & 3 or 4, multiple lines & \citet{Morris:2015}
\enddata
\tablecomments{Table describing spectroscopic redshift compilation. Only robust redshifts are used, with multiple lines identifying the redshifts.}

\tablenotetext{a}{VVDS stands for VIMOS VLT Deep Survey, GRAPES for Grism ACS Program for Extragalactic Science, \citep{Pirzkal:2004}, GOODS for Great Observatories Origins Deep Survey, GMASS for Galaxy Mass Assembly ultra-deep Spectroscopic Survey, and 3D-HST is a large HST program \citep{Brammer:2012}. }
\tablenotetext{b}{Either Very Large Telescope (VLT) or Hubble Space Telescope (HST).}
\tablenotetext{c}{Either FOcal Reducer and low dispersion Spectrograph (FORS), VIsible Multi-Object Spectrograph (VIMOS), Advanced Camera for Surveys (ACS), or Wide Field Camera 3 (WFC3).}
\end{deluxetable*}

\section{Photometric Redshifts}
\label{photoz}

Photometric redshifts are an accepted and robust procedure to estimate redshifts of galaxies without spectroscopic information \citep[e.g.,][]{Koo:1985, Benitez:2000, Coe:2006, Ilbert:2006, Ilbert:2009, Hildebrandt:2010, Dahlen:2013}. There are a large number of photometric redshift software packages available, and the differences between them are summarized by \citet{Hildebrandt:2010} and \citet{Dahlen:2013}. In addition to differences in the methodologies of the code, the choice of templates and filters affect the final redshift estimates. For these reasons, we present and compare redshifts from two of the best-performing packages with different methodologies, priors and template sets. 

\begin{figure}[b!]
\center{
\includegraphics[scale=0.4, viewport=10 10 550 420,clip]{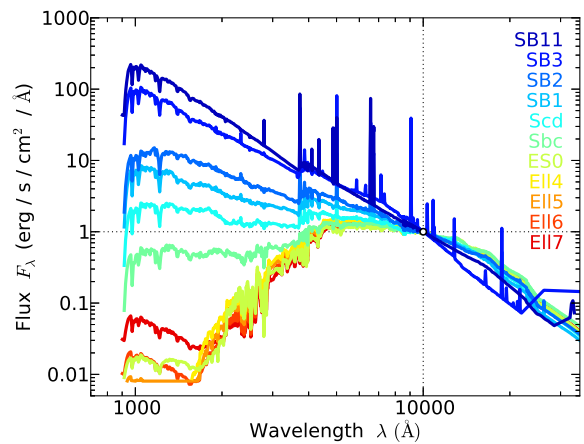}
}
\caption{BPZ templates including four elliptical galaxies (Ell), one Lenticular (ESO), two spirals (Sbc and Scd), and four starbursts (SB). These templates are based on those from PEGASE \citep{Fioc:1997} but re-calibrated based on observed photometry and spectroscopic redshifts from FIREWORKS \citep{Wuyts:2008}.
\label{fig:templates} 
}
\end{figure}

The Bayesian Photometric Redshift (BPZ) algorithm \citep{Benitez:2000, Benitez:2004, Coe:2006} is the primary redshift software used due to its robust performance \citep{Hildebrandt:2010}, familiarity of the code to the authors, and best performance as detailed in Section \ref{quality}. The BPZ SED templates and priors have been significantly updated (Benitez et al. in prep.) as described in \citet{Coe:2013} and \citet{Benitez:2014}. Specifically, the model spectral energy distributions (SEDs) are based on those from PEGASE \citep{Fioc:1997} but re-calibrated based on observed photometry and spectroscopic redshifts from FIREWORKS \citep{Wuyts:2008}. The templates are shown in Figure 10 of \citet{Benitez:2014}, and in Figure \ref{fig:templates} here. These templates include four elliptical galaxies, one Lenticular, two spirals, and four starbursts. The templates include emission lines, and we interpolate between each pair of adjacent templates to create nine intermediate templates, yielding 111 possibilities in BPZ. The Bayesian prior is based on luminosity functions observed in COSMOS \citep{Ilbert:2009}, GOODS-MUSIC \citep{Grazian:2006, Santini:2009}, and the UDF \citep{Coe:2006}. The software is also modified from the publicly released version with minor improvements, such as matching the output uncertainties to the observed scatter. 

In addition to the BPZ redshifts, we also determine a second set of redshifts using the EAZY software \citep{Brammer:2008}, another top performer by the \citet{Hildebrandt:2010} study. This code not only provides another independent methodology including the use of a template error function \citep{Brammer:2008}, but also uses a different set of galaxy templates including emission lines based on star formation rates \citep{Brammer:2011}, and the inclusion of a very dusty template (Brammer et al. in prep). 

The quality of the photometric redshift fit is described by two quantities, {\tt ODDS} and reduced chi square. The {\tt ODDS} parameter measures the spread in the probability distribution function, $P(z)$, representing the probability of a galaxy being at a specific redshift. The ODDS for BPZ is defined as the integrated $P(z)$ contained within $2*0.03*(1+z)$. When $P(z)$ has multiple peaks, the value of {\tt ODDS} is low and the resulting redshift uncertainties are large. If $P(z)$ has only a single peak, then the {\tt ODDS} describes the width of the distribution, with a maximum value of ODDS of 1.0. Typically {\tt ODDS}$>0.9$ is considered to be a high-probability determination of the redshift \citep{Benitez:2000, Benitez:2004,Coe:2006}. 

The second quantity is a reduced chi square goodness of fit. In addition to the regular chi square used in the calculations, BPZ also reports a modified reduced chi square, $\chi ^2_{\mathrm{mod}}$. The $\chi ^2_{\mathrm{mod}}$ is similar to a normal reduced chi square, but includes an additional uncertainty for the SED templates in addition to the uncertainty in the galaxy photometry \citep{Coe:2006}. The added uncertainty for the SED was determined such that the resultant $\chi ^2_{\mathrm{mod}}$  is a more realistic measure of the quality of the fit \citep[for more discussion of $\chi ^2_{\mathrm{mod}}$, see][]{Rafelski:2009}. For the EAZY algorithm, a normal reduced chi square, $\chi ^2_\nu$, is used. While strictly speaking a $\chi ^2_{\nu}\sim1$ would be considered a good fit, in practice a $\chi ^2_{\mathrm{mod}}<4$ was found to indicate a relatively good fit with robust redshifts \citep{Rafelski:2009}. 

A comparison of $\chi ^2_{\mathrm{mod}}$ from BPZ and $\chi ^2_\nu$ from EAZY finds that they differ as a function of magnitude, with $\chi ^2_\nu$ being larger. For fainter galaxies (F606W magnitude $>27$), they differ by a factor of $\sim2$. However, as the brightness of the galaxies increases, the ratio increases as does the scatter. For bright galaxies (F606W magnitude $<25$), the ratio increases to $\sim11$. This is reasonable, given that  $\chi ^2_{\mathrm{mod}}$  was specifically designed to correct for that fact that galaxies with small photometric uncertainties tend to have high $\chi ^2_\nu$ even for good fits \citep{Coe:2006, Rafelski:2009}, since systematic uncertainties for the SED templates are not taken into account. In order to have a similar number of targets selected as having good $\chi ^2$ for both BPZ and EAZY in the investigations below, a $\chi ^2_{\nu}<10$ is used as an indicator of a good fit for EAZY redshifts in this paper. Restrictions to lower $\chi ^2$ and higher {\tt ODDS} will improve the redshift quality further, yet at the cost of sample size.

While naively one may expect that the inclusion of mid-IR data may improve the photometric redshifts, empirical tests show that many algorithms (including BPZ) actually perform worse when including the mid-IR photometry \citep{Hildebrandt:2010}. This was confirmed in tests on our data set as well, and is likely due to insufficient knowledge of the template SEDs in the mid-IR, due to degeneracies in the dust and PAH spectral features. In addition to such mid-IR data having significantly worse PSF FWHM's than the data described here which complicates its use, and the data being significantly less deep than the other data sets used here, there is also no improvement in the photometric redshifts with their inclusion. Therefore, such data are not included for the photometric redshift determinations. 

The catalog and photometric redshifts are not optimized for searches of $z\gtrsim8$ galaxies, due to choice made in the catalog creation to optimize intermediate redshift source selection. First, as described in Section \ref{detect_image}, the detection image consists of 8 images including the optical data. Hence, faint galaxies appearing only in the reddest NIR bands may not be detected at all, as the NIR flux is averaged with that in bluer bands in which the galaxy's Lyman break flux decrement causes the galaxy to drop out. Furthermore, as discussed in Secton \ref{catalog}, single band detections are excluded from the catalog to avoid spurious detections, and hence the highest redshift galaxies that only appear in the F160W image are excluded. While we find no new $z\gtrsim8$ galaxies, we do recover candidates out to $z\sim8.3$ and NIR mag $\sim29$ previously published in the UDF \citep{Ellis:2013, McLure:2013, Schenker:2013, Oesch:2013}.

\begin{deluxetable*}{ccccccccccc}
\tabletypesize{\scriptsize}
\tablecaption{Reliable Spectroscopic Redshifts of Galaxies in the UDF
\label{tab:specz}}
\tablewidth{0pt}
\tablehead{
\colhead{ID} &
\colhead{RA} &
\colhead{DEC} &
\colhead{$z_{\mathrm{spec}}$} &
\colhead{$z_{\mathrm{BPZ}}$\tablenotemark{a}} &
\colhead{$\chi^2_{\mathrm{mod}}$\tablenotemark{b}} & 
\colhead{$\tt ODDS$\tablenotemark{c}} &
\colhead{$z_{\mathrm{EAZY}}$\tablenotemark{d}} &
\colhead{$\chi^2_\nu$\tablenotemark{e}} & 
\colhead{$\tt ODDS$\tablenotemark{f}} &
\colhead{Reference\tablenotemark{g}}}
\startdata
373 &53.15437485 &-27.82148101 &1.14 &1.22$_{-0.10}^{+0.10}$ &0.51 &0.99 &1.11$_{-0.10}^{+0.12}$ &1.22 &1.00 &5 \\
534 &53.16170093 &-27.81925383 &0.67 &0.65$_{-0.08}^{+0.06}$ &0.10 &1.00 &0.61$_{-0.11}^{+0.09}$ &0.56 &1.00 &6 \\
865 &53.17450843 &-27.81495550 &0.67 &0.64$_{-0.08}^{+0.07}$ &0.06 &0.99 &0.61$_{-0.10}^{+0.09}$ &0.84 &1.00 &1 \\
983 &53.14989262 &-27.81400031 &1.31 &1.28$_{-0.10}^{+0.10}$ &0.22 &0.99 &1.24$_{-0.11}^{+0.11}$ &1.44 &1.00 &6 \\
1035 &53.17634676 &-27.81475285 &2.44 &2.42$_{-0.15}^{+0.15}$ &0.68 &1.00 &1.99$_{-1.94}^{+0.13}$ &7.85 &0.93 &6 \\
1060 &53.15915906 &-27.81374038 &1.77 &1.74$_{-0.13}^{+0.11}$ &0.27 &0.99 &1.71$_{-0.08}^{+0.08}$ &2.45 &1.00 &8 \\
1077 &53.16529350 &-27.81405511 &3.06 &3.32$_{-0.19}^{+0.18}$ &0.73 &1.00 &3.04$_{-0.17}^{+0.29}$ &1.03 &1.00 &2 \\
1134 &53.16817307 &-27.81293086 &0.96 &1.00$_{-0.09}^{+0.09}$ &0.18 &0.99 &0.79$_{-0.10}^{+0.09}$ &1.48 &1.00 &6 \\
1220 &53.17932560 &-27.81252367 &1.77 &1.83$_{-0.13}^{+0.12}$ &0.89 &0.99 &1.74$_{-0.06}^{+0.06}$ &11.48 &1.00 &9 \\
1438 &53.14508947 &-27.80985318 &1.25 &1.26$_{-0.11}^{+0.09}$ &0.01 &0.99 &1.21$_{-0.11}^{+0.10}$ &0.93 &1.00 &9
\enddata
\tablecomments{Table \ref{tab:specz} is published in its entirety in a machine-readable form in the online version of the Astronomical Journal. 
A portion is shown here for guidance regarding its form and content.  }
\tablenotetext{a}{Bayesian Photometric Redshift (BPZ)  and uncertainty from 95\% confidence interval.}
\tablenotetext{b}{Modified reduced chi square fit, where the templates are given uncertainties.}
\tablenotetext{c}{Integrated $P(z)$ contained within $2*0.03(1+z_{\mathrm{BPZ}})$.}
\tablenotetext{d}{EAZY redshift and uncertainty from 95\% confidence interval.}
\tablenotetext{e}{Reduced chi square fit.}
\tablenotetext{f}{Integrated $P(z)$ contained within $0.2(1+z_{\mathrm{EAZY}})$.}
\tablenotetext{g}{(1) \citet{LeFevre:2004} (2) \citet{Szokoly:2004} (3) \citet{Mignoli:2005} (4)\citet{Daddi:2005ek} (5) \citet{Vanzella:2005, Vanzella:2006, Vanzella:2008, Vanzella:2009} (6) \citet{Popesso:2009} (7) \citet{Balestra:2010} (8) \citet{Kurk:2013} (9) \citet{Morris:2015}}
\end{deluxetable*}

\subsection{Sample of Spectroscopic Redshifts}
\label{specz}
In order to assess the quality of the photometric redshifts, a large sample of high-quality spectroscopic redshifts is needed. We compiled a total of 176 robust spectroscopic redshifts from 9 separate surveys, which almost doubles the compilation from \citet{Rafelski:2009}. Almost half the resultant sample have more than one redshift measurement, and three of the measurements are inconsistent. For those three, the best redshift is selected based on the quality of the original data. While the number of spectroscopic redshifts could be significantly increased by including lower-quality redshifts, doing so would introduce an uncertainty when photometric and spectroscopic redshift measurements do not agree, making it unclear which redshift is incorrect.

The source of the spectroscopic redshifts are outlined in Table \ref{tab:specz_comp}, along with the telescopes and instruments used, the number of sources from each survey, and the quality cut applied. In general, the quality cut requires redshifts to be identified with multiple lines and good SN. The 3D-HST redshifts used differ from the official 3D-HST grism redshifts, which include photometry in their redshift determinations (see Section \ref{grism}), and are from and independent analysis by \citet{Morris:2015}. None of the redshifts used include photometry to determine the redshift, so they are an independent check on the photometric redshifts. Of the 176 spectroscopic redshifts described above, 7 are stars and are not included in the investigations below.

\begin{figure}[t!]
\center{
\includegraphics[scale=0.4, viewport=5 10 550 360,clip]{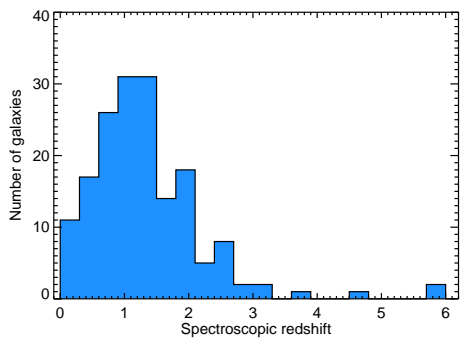}
}
\caption{ \label{fig:speczhist} 
Spectroscopic redshift distribution of 169 galaxies in the Hubble Ultra Deep Field. Most of the spectroscopic redshifts are at $z<2$. 
}
\end{figure}

The 169 reliable spectroscopic redshifts of galaxies are listed in Table \ref{tab:specz}. The table includes the main catalog ID number, spectroscopic redshift, photometric redshift, {\tt ODDS}, and $\chi ^2_{\mathrm{mod}}$. Although not included in this table, the photometric redshifts of the 7 stars are incorrect because a stellar template is not included in the photometric fits, and are all identified in the star catalog \citep{Pirzkal:2005}. Figure \ref{fig:speczhist} shows a histogram of the galaxy redshifts, and shows that the vast majority of spectroscopic redshifts are at $z<2$, with a peak at $z\sim1$. This means that the photometric redshifts can be well tested at $z\sim1$, but that redshifts at $z>2$ will be less well vetted. Some improvement on the redshift distribution is obtained with the grism redshifts, although with their own caveats as discussed in Section \ref{grism}.

Neither of the redshift codes is explicitly tuned or trained using these spectroscopic redshifts, although the templates used by BPZ are re-calibrated based on observed photometry and spectroscopic redshifts from the FIREWORKS catalog \citep{Wuyts:2008}. This catalog includes about half of the redshifts presented here from spectroscopic campaigns before 2008. This does not significantly affect our quality checks below, as there are over 6000 redshifts in the FIREWORKS catalog which are used in refining the templates. Moreover, the photometry used would not include the new NUV or NIR data in the UDF, which are primarily responsible for the improvements in the redshifts derived here. Lastly, the improvements in redshift derivations with the addition of the NUV and NIR data discussed in Section \ref{uvimprove} are compared to redshifts determined with the same templates, and thus the comparisons are unbiased.

\subsection{Quality of Photometric Redshifts}
\label{quality}
\begin{figure}[t!]
\center{
\includegraphics[scale=0.4, viewport=20 10 480 550,clip]{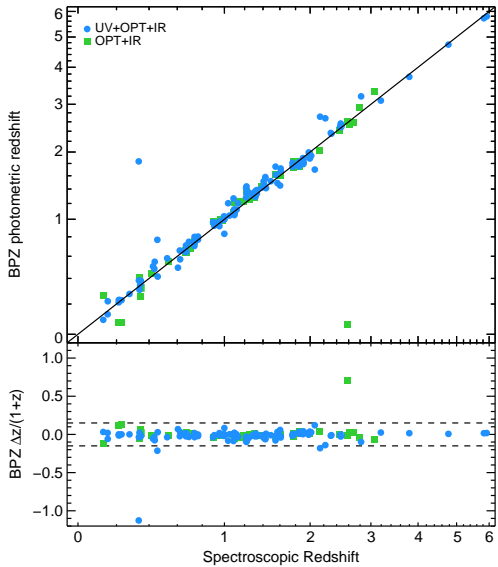}
}
\caption{ \label{fig:photspecbpz} 
Photometric versus spectroscopic redshifts for 169 galaxies in the UDF on a `pseudolog' scale, showing the high quality of the BPZ photometric redshifts. The blue circles represent redshifts for galaxies that are covered by the NUV data, and the green squares are for redshifts without NUV data. All redshift determinations include NIR and optical data. Filled symbols represent photometric redshifts with good {\tt ODDS} and chi square, {\tt ODDS} $>0.9$ and $\chi ^2_{\mathrm{mod}}<4$ for BPZ and {\tt ODDS} $>0.9$ and $\chi ^2_{\nu}<10$ for EAZY. The open symbols on the other hand are for redshifts that may have multiple peaks in $P(z)$ or may have a low quality fit, {\tt ODDS} $<0.9$ or $\chi ^2_{\mathrm{mod}}>4$ for BPZ and  {\tt ODDS} $<0.9$ or $\chi ^2_{\nu}>10$ for EAZY. Dashed lines represent outlier cutoff, defined as $\left| \Delta z \right| / (1+z_{\mathrm spec}) < 0.15$. 
}
\end{figure}

\begin{figure}[t!]
\center{
\includegraphics[scale=0.4, viewport=20 10 480 550,clip]{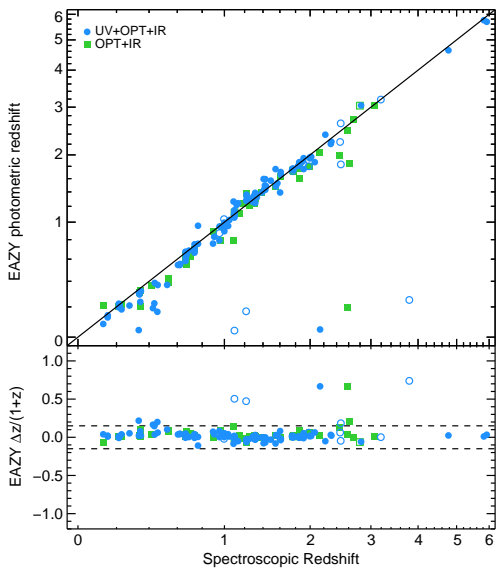}
}
\caption{ \label{fig:photspeceazy} 
This figure is the same as Figure \ref{fig:photspecbpz}, except for the EAZY photometric redshifts instead of the BPZ redshifts.
}
\end{figure}

The quality of the photometric redshifts is evaluated by a comparison of the subset of galaxies with high quality spectroscopic redshifts, such that it is extremely likely that any discrepancies are due to inaccuracies of the photometric rather than the spectroscopic redshifts. Figures \ref{fig:photspecbpz} and \ref{fig:photspeceazy} show the very good quality of the photometric redshifts in the UDF, with very few outliers and a tight relation around the one-to-one line. The comparison plot is on a `pseudolog' (log($1+z$)) scale to spread out the data points at lower redshifts. No optimizations are performed for this comparison, so it is representative of the general quality of the photometric redshifts, although the galaxies with spectroscopic redshifts consist of the brighter galaxies in the UDF. The quality of the redshifts are quantified by the scatter in the difference between the photometric and spectroscopic redshifts, and by the fraction of outliers far from the line.

\subsubsection{Scatter of Redshifts}
\label{scatter}

The scatter in the difference of the photometric and spectroscopic redshifts provides a metric to quantify the general accuracy of the photometric redshifts. 
To minimize the impact of outliers on the measurement of the scatter, the normalized median absolute deviation \citep[$\sigma_{\rm NMAD}$;][]{Brammer:2008} is used to define the scatter:
\begin{equation}
\sigma_{\rm NMAD} = 1.48 \times \mathrm{median} \left| \frac{\Delta z-\mathrm{median}(\Delta z)}{1+z_{\mathrm spec}} \right|, 
\end{equation}
\noindent where $\Delta z = z_{\rm spec}-z_{\rm phot}$,  $z_{\rm spec}$ is the spectroscopic redshift, and $z_{\rm phot}$ is the photometric redshift. The $\sigma_{\rm NMAD}$ is equivalent to the standard deviation for a Gaussian distribution, but is less sensitive to outliers. There is a slightly different version of $\sigma_{\rm NMAD}$ also used in the literature \citep[e.g.,][]{Ilbert:2009, Dahlen:2013}, but the above version is more robust to outliers due to the use of the second median.

The $\sigma_{\rm NMAD}$ is very small for our sample, significantly improving on previous photometric redshifts in the UDF, with $\sigma_{\rm NMAD}=$ 0.028 for BPZ and $\sigma_{\rm NMAD}=$ 0.030 for EAZY.  These numbers drop even further for BPZ when considering only the sample with NUV data, with $\sigma_{\rm NMAD}=$ 0.026, while there is an increase for EAZY, with $\sigma_{\rm NMAD}=$ 0.035 (see Section \ref{photoz}).

The last full UDF photometric redshift catalog by \citet{Coe:2006} has a $\sigma_{\rm NMAD}= $0.068 using the same spectroscopic sample, and therefore the redshifts presented here are a factor of two better than before, based on this metric. Another catalog of photometric redshifts in the UDF is presented by \citet{Cameron:2011}, although it only includes the 1052 galaxies brighter than mag$_{\rm F160W}<27$ that are in the smaller deep NIR region. Matching this sample to the spectroscopic redshifts defined above yields $\sigma_{\rm NMAD}=$0.042 based on 93 redshifts. 

In addition to the UDF redshift catalogs, two recent catalogs of GOODS-South encompass the UDF, and have photometric redshifts for all 168 spectroscopic redshifts. The CANDELS team calculated redshifts by combining multiple redshifts from different software packages, using 14 bandpasses including ground-based u-band data and Spitzer data, and have a $\sigma_{\rm NMAD}= $0.022 (Santini et al. 2015, in prep). The 3D-HST team also calculated photometric redshifts using EAZY, using up to 40 bandpasses including significant numbers of ground-based data and Spitzer data, and have a $\sigma_{\rm NMAD}= $0.013 \citep{Skelton:2014}. While these new catalogs have better $\sigma_{\rm NMAD}$ than our catalog, our results have about a factor of two improvement in outlier fraction (see Section \ref{outlier}) below. 
Our $\sigma_{\rm NMAD}$ also compares well to other larger studies of different fields with many bandpasses, such as the COSMOS redshifts \citep[$\sigma_{\rm NMAD}=0.007-0.033$;][]{Ilbert:2009} or the CANDELS GOODS-south redshifts \citep[$\sigma_{\rm NMAD}=0.03-0.1$;][]{Dahlen:2013}, although our galaxy sample is fainter. 

\subsubsection{Outlier Fractions}
\label{outlier}

Photometric redshifts can have ``catastrophic'' redshift errors, caused by template mismatches, when the wrong redshift is assigned to a galaxy, and the probability distribution function of the photometric redshift does not include the true redshift \citep{Ellis:1997, FernandezSoto:1999, Benitez:2000, Rafelski:2009}. Most of the time the probability function correctly includes both redshift possibilities, but if there is a single peak at the wrong redshift with an acceptable chi square value, then the galaxy is unambiguously assigned to the wrong redshift. The rate of these catastrophic errors in the UDF has significantly decreased over previous work \citep[e.g.,][]{Coe:2006, Rafelski:2009}, but still exists for various reasons.

Any individual galaxy redshift is only as good as its object definition, and bad object definitions will result in poor photometric redshifts. For instance, if multiple galaxies are included in a single aperture due to insufficient deblending, then the resultant photometric redshift will be incorrect. While every attempt is made to avoid this in the catalog, one such example is very evident in the comparison in the top left panel of Figure \ref{fig:photspecbpz}. Object 10157 is composed of two overlapping galaxies at high and low redshifts, which results in an incorrect photometric redshift at $z_{\rm phot}\sim2$ in the Figure.  While the cause of the incorrect photometric redshift is understood, it and others like it are still included in all subsequent analysis to avoid any selection biases. A random selection of galaxies using photometric redshifts would likewise also include such object definitions. Therefore, including these galaxies in the comparisons below makes the outlier fractions consistent with what a user of the catalog will experience. For small samples of galaxies, users may wish to manually inspect the segmentation map and the photometric redshift fits, which would yield a more robust sample. Galaxy cutouts and SED plots are available on the web at: \url{http://www.rafelski.com/uvudf/catalogs.html} for the BPZ redshifts.

There are multiple methods to define an outlier in a comparison of spectroscopic and photometric redshifts. One method is to consider objects at  $>5\sigma_{\rm NMAD}$ as outliers, since these redshifts are significantly more deviant than the scatter \citep{Brammer:2008} . However, in that case the photometric redshifts with a larger scatter and the same number of significantly deviant redshifts may have a lower outlier fraction than a dataset with a smaller scatter. An alternative method is compare the outliers to an absolute deviation, such as $\left| \Delta z \right| / (1+z_{\mathrm spec}) > 0.15$ \citep{Ilbert:2006, Ilbert:2009, Hildebrandt:2010, Brammer:2011, Dahlen:2013}, although this could result in false outliers if the scatter is large. Given the $\sigma_{\rm NMAD}$ of our sample, these two methods are almost identical for our photometric redshifts, and the second method is used for the outlier fraction (OLF) discussed below. 

There are a total of 4 outliers with $\left| \Delta z \right| / (1+z_{\mathrm spec}) > 0.15$ for BPZ redshifts, and 10 outliers for the EAZY redshifts. These correspond to OLFs  of 2.4\% and 5.9\% respectively.  If limiting the spectroscopic sample to those with with good {\tt ODDS} and $\chi ^2$ as defined in Section \ref{photoz}, then the outlier fractions change to 2.4\% and 3.9\% for BPZ and EAZY respectively. The OLF for EAZY drops to 2.7\% when considering only the sample with NUV data (three outliers), while the BPZ fraction stays the same at 2.4\% (although with one fewer outlier). The OLFs can be made even smaller if a more stringent cutoff for $\chi ^2$ is used, although that also reduces the size of the sample.

This catalog is a significant improvement over the previous catalogs of the UDF, which have an OLF of 16.4\% \citep{Coe:2006} and 4.3\% \citep{Cameron:2011} when determined with the new larger spectroscopic sample defined above. Note that the OLF of \citet{Coe:2006} decreases to 10.6\% when only including galaxies with good {\tt ODDS} and $\chi ^2$. Our redshifts thereby provide a factor of three improvement in outlier rate than the previous full redshift catalog of the UDF, and a factor of two better than those of \citep{Cameron:2011}.

The catalog also compares well to the GOODS-South catalogs, which have outlier fractions of 3.0\% and 5.3\% for CANDELS and 3D-HST respectively when comparing to the same 168 spectroscopic redshifts. When comparing to the smaller subsample that also have NUV data, both the CANDELS and 3D-HST outlier fractions stay about the same at 3.3\% and 5.6\%, showing no improvement in this subsample of spectroscopic redshifts. This suggests that it is indeed the addition of the NUV data, and not sample selection, that is responsible for the improved outlier fractions in the NUV sample above. 


\subsubsection{Combining Multiple Redshift Methods}

These outlier galaxies are not the same in the two photometric redshift samples, and thus outliers can be further avoided by considering only galaxies with similar photometric redshifts between the two software packages, defined as $\left| \Delta z_{phot} \right| / (1+z_{\mathrm BPZ}) < 0.15$, where $\Delta z_{phot}=z_{\rm BPZ}-z_{\rm EAZY}$. The sample that consists of matching photometric redshifts yields an OLF of 0.63\%, which consists of a single outlier. Restricting the sample to good {\tt ODDS} and $\chi ^2$ for both redshift codes keeps the single outlier, with an OLF of 0.68\%.
Requiring matching redshifts between codes reduces the sample size to 93\% and 86\% for the two respectively.

The matched photometric redshift sample does not significantly improve the scatter $\sigma_{\rm NMAD}$, with $\sigma_{\rm NMAD} = 0.027$ for both BPZ and EAZY when restricting to matched photometric redshifts. On the other hand, \citet{Dahlen:2013} found that combining multiple redshifts results yielded the best photometric redshifts. Taking an average photometric redshift from the two codes improves  $\sigma_{\rm NMAD}$ for the sample with good {\tt ODDS} and $\chi ^2$ for both methods, with $\sigma_{\rm NMAD} = 0.026$ for the average redshift value. 

Since the improvements of combining both codes helps improve outlier fractions, and this is another option for the reader if they so choose. However, for simplicity the rest of the paper examines only a single redshift code. While both redshift codes provide similar quality redshifts, the BPZ redshifts are utilized henceforth due to slightly better $\sigma_{\rm NMAD}$ and OLF (see also Section \ref{grism}), and the authors familiarity with the code. The catalog in Section \ref{catalog} presents both redshifts, and hence the reader can choose which redshifts to use.

\subsection{Comparison with Grism Redshifts}
\label{grism}

The use of the WFC3 grisms on HST to determine redshifts is increasing at a rapid pace \citep[e.g.,][]{Brammer:2012, Colbert:2013, Atek:2014}, providing large numbers of galaxy redshifts. The UDF is covered by the G141 WFC3 grism by both 3D-HST and CANDELS, and a special data release with redshifts \citep{Brammer:2012,Brammer:2013} for 228 galaxies in the UDF is available. While the spectroscopic sample does include 28 redshifts from these data in the spectroscopic sample when based on two emission lines with good signal to noise \citep{Morris:2015}, the other 200 galaxy redshifts are not included because the 3D-HST redshifts are determined from the combination of the grism data and the photometric data. In essence this is a combination of grism emission or absorption lines and slope combined with the simultaneous determination of a photometric redshift with EAZY. While a pure photometric redshift from EAZY does not always agree with these redshifts, they are still based on the photometry and in such are not entirely independent tests for the photometric redshifts. In addition, while spectroscopic redshifts typically have a metric to select the more robust redshifts, the grism redshifts are presented without a quality flag. Therefore, we do not combine these redshifts with the robust spectroscopic sample defined in Section \ref{specz}.

\begin{figure}[t!]
\center{
\includegraphics[scale=0.4, viewport=5 10 550 360,clip]{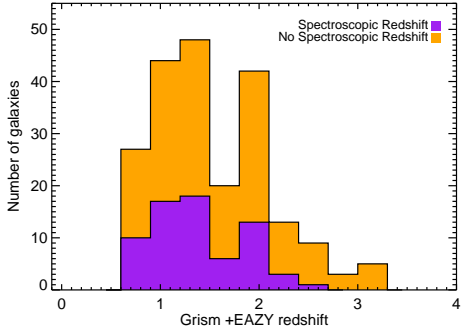}
}
\caption{ \label{fig:grismhist} 
Redshift distribution of 214 grism redshifts. The orange represents all grism redshifts not covered by the spectroscopic redshifts, and the purple represents the 68 grism redshifts for which spectroscopic redshifts are available.
}
\end{figure}

While the 3D-HST grism redshifts are not independent of the photometry, they do provide another avenue to test the photometric redshifts. Since there are no quality flags, the redshifts were manually inspected, and 214 of the 228 redshifts were selected as robust (G. Brammer, private communication). Of these, 143 do not have robust spectroscopic redshifts and are a new test of the photometric redshifts, and the distribution of the grism redshifts are shown in Figure \ref{fig:grismhist}. This includes 40 redshifts at $z>2$, almost twice the sample size of the 23 spectroscopic redshifts at $z>2$.

\begin{figure}[t!]
\center{
\includegraphics[scale=0.4, viewport=20 10 480 550,clip]{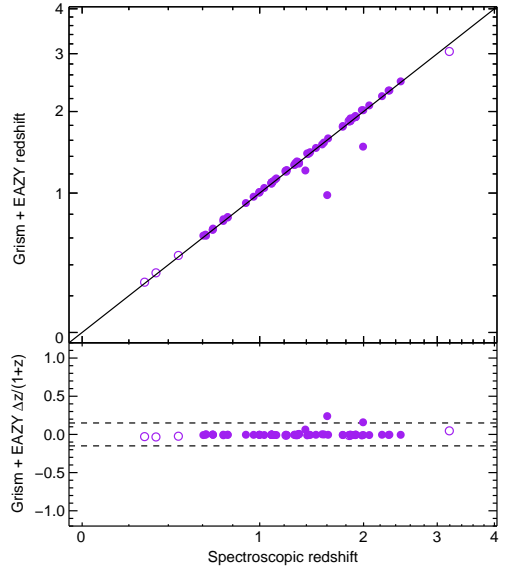}
}
\caption{ \label{fig:grismcheck} 
Grism versus photometric redshifts for the 71 galaxies in the UDF with both grism and spectroscopic redshifts on a `pseudolog' scale (solid purple circles). An additional 6 galaxies are shown as open purple circles, which are from the sample of 14 grism redshifts excluded in Section \ref{grism} as potentially not being robust, and also have a spectroscopic redshift. The grism and spectroscopic redshifts are in good agreement except for three galaxies, showing that the grism redshifts are pretty good, although not as good the sample of robust spectroscopic redshifts described in Section \ref{specz}.}
\end{figure}

\begin{figure}[h!]
\center{
\includegraphics[scale=0.4, viewport=20 10 480 550,clip]{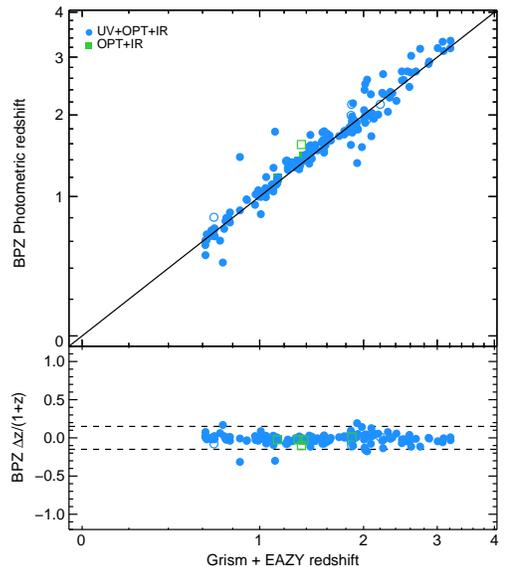}
}
\caption{ \label{fig:grism} 
Photometric versus grism redshifts for 211 galaxies in the UDF on a `pseudolog' scale, showing the high quality of the photometric redshifts (excluding the three disagreements from the left panel plots). The blue circles represent redshifts for galaxies that are covered by the NUV data, and the green squares are for redshifts without NUV data. All redshift determinations include NIR and optical data. Filled symbols represent photometric redshifts with good {\tt ODDS} and $\chi ^2$, {\tt ODDS} $>0.9$ and $\chi ^2_{\mathrm{mod}}<4$. The open symbols on the other hand are for redshifts that may have multiple peaks in $P(z)$ or may have a low quality fit, {\tt ODDS} $<0.9$ or $\chi ^2_{\mathrm{mod}}>4$.
Dashed lines represent outlier cutoff, defined as $\left| \Delta z \right| / (1+z_{\mathrm spec}) < 0.15$.}
\end{figure}

As a test of the grism redshifts, the sub-sample of redshifts that are also in our sample of robust spectroscopic redshifts are compared in Figure \ref{fig:grismcheck}. This shows that of the 71 galaxies with both grism and spectroscopic redshifts, three do not agree, and two of these would be an outlier by the definition used for the photometric redshift comparison, $\left| \Delta z_{\mathrm grism} \right| / (1+z_{\mathrm spec}) > 0.15$. These two outlier redshifts are strongly influenced by a single emission line, while the other show no emissions lines at all, but the redshift is based on the SED and weak absorption lines.  While it is possible that the spectroscopic redshifts are in error, other work also find that the grism redshifts sometimes disagree with the spectroscopic redshifts \citep{Kriek:2014}. Additionally, the photometric redshifts agree with the spectroscopic redshifts, suggesting it is more likely that the grism redshifts are in error. Regardless, the agreement in Figure \ref{fig:grismcheck} is sufficiently good that it makes sense to compare the grism redshifts with the photometric redshifts, assuming that the grism redshifts are correct. For the rest of the comparison with the grism redshifts, the three grism redshifts not in agreement with the spectroscopic redshifts are dropped from the analysis (e.g., Figure \ref{fig:grism}). 

The grism redshifts are compared to the photometric redshifts in Figure \ref{fig:grism}. The scatter is similar to that when comparing with the spectroscopic redshifts, $\sigma_{\rm NMAD}=$ 0.031 and $\sigma_{\rm NMAD}=$ 0.040 for BPZ and EAZY.  The OLFs are 2.9\% and 4.7\% for BPZ and EAZY, which become 2.5\% and 4.9\% respectively for good {\tt ODDS} and $\chi ^2$.  The $\sigma_{\rm NMAD}$ and OLFs remain the same when considering the sample with NUV data, as most of the grism redshifts fall in the area covered by NUV data. 

Similar to the spectroscopic sample, this is an improvement over the previous catalogs when compared to the grism redshifts. The \citet{Coe:2006} redshifts have $\sigma_{\rm NMAD}=$ 0.031 and OLF of 15\%  (12\% with good {\tt ODDS} and $\chi ^2$), and the catalog by \cite{Cameron:2011} have $\sigma_{\rm NMAD}=$ 0.059 and an OLF of 6.8\%. This confirms the finding from the spectroscopic sample that the photometric redshifts presented here are  robust, with significant improvements over previous catalogs. There are also clear improvements when including the NUV data, especially in the OLFs as discussed below.

\begin{figure*}[h!]
\center{
\includegraphics[scale=0.4, viewport=5 10 550 360,clip]{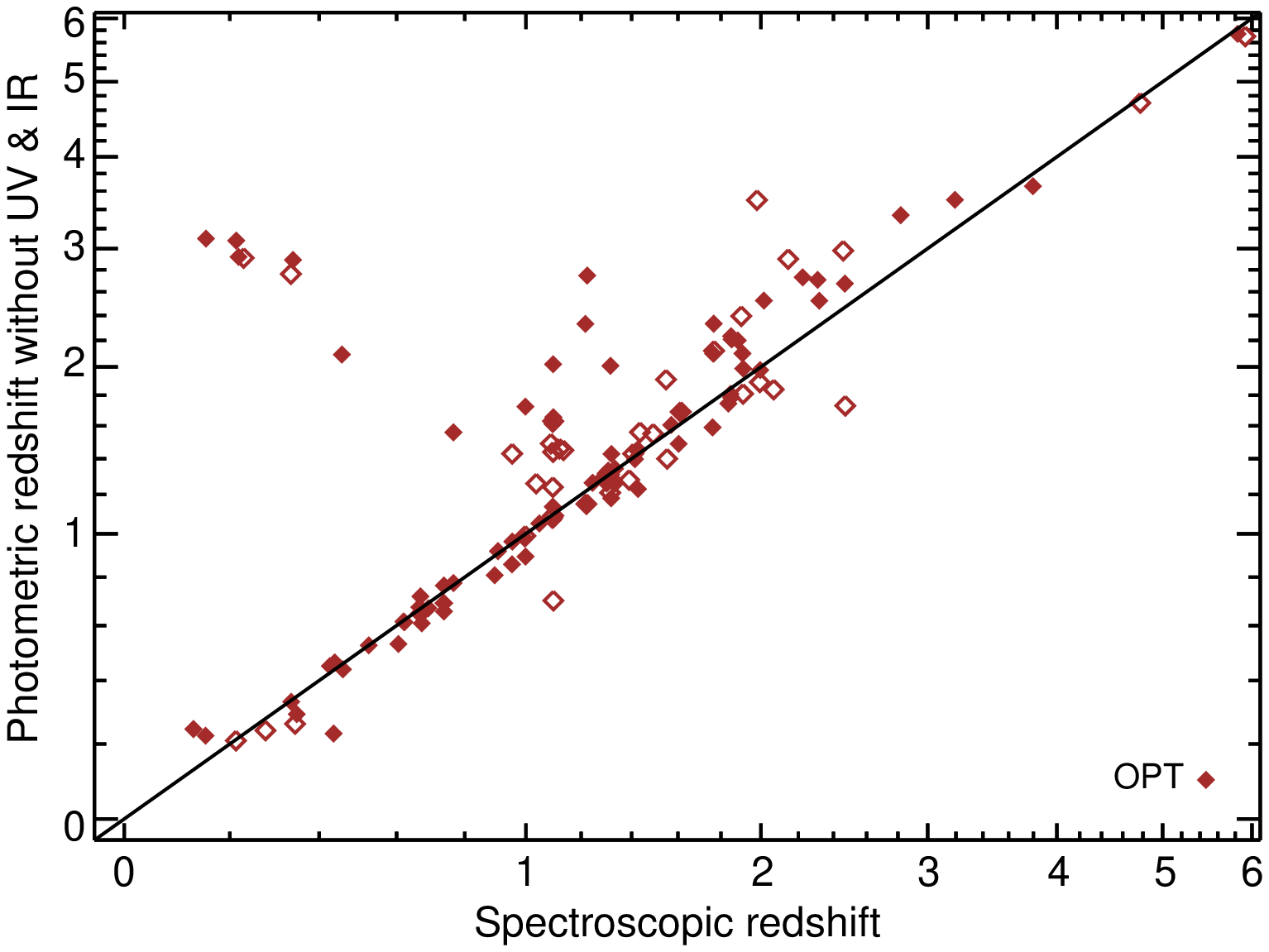}
\includegraphics[scale=0.4, viewport=5 10 550 360,clip]{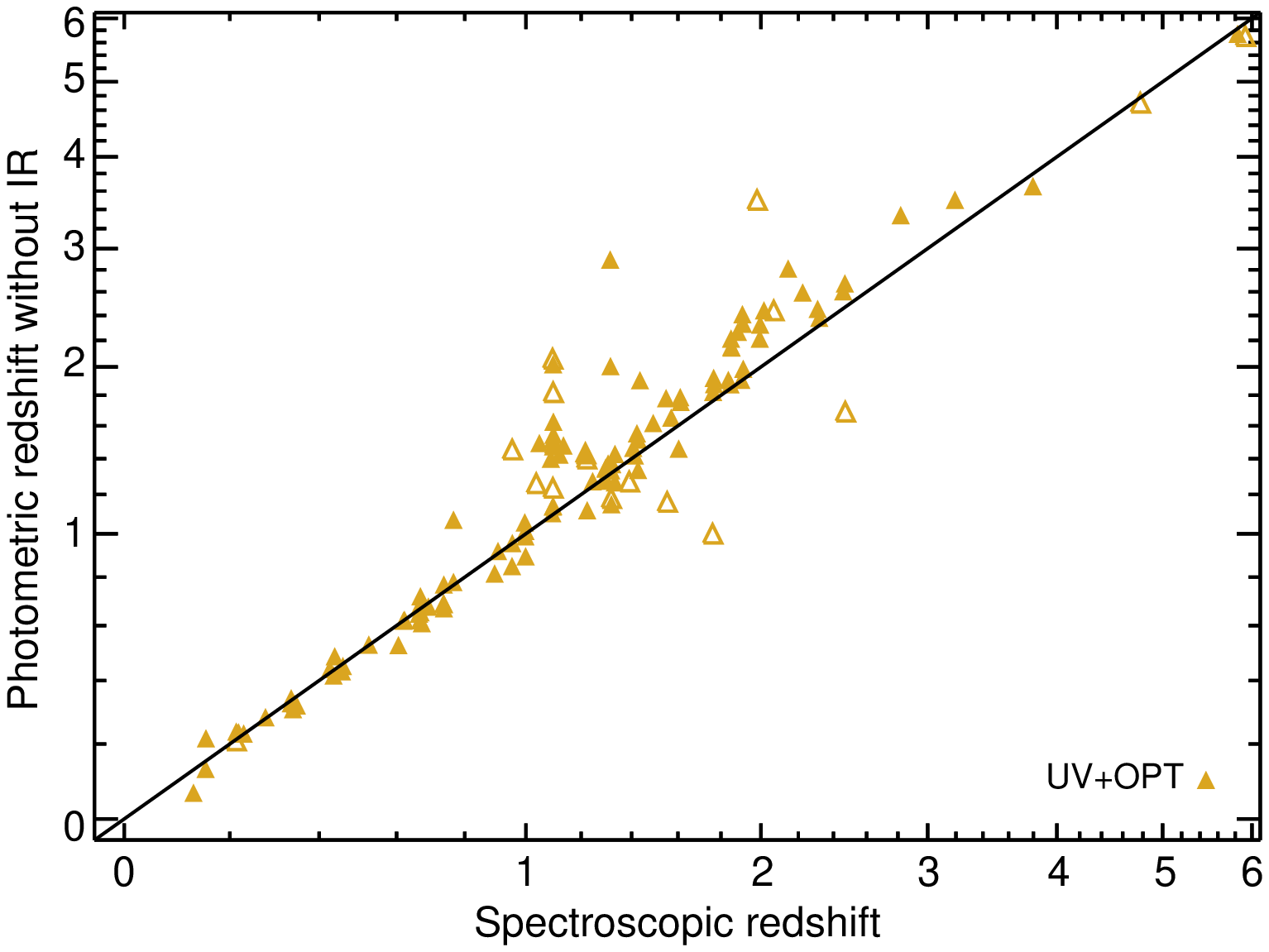}
\includegraphics[scale=0.4, viewport=5 10 550 360,clip]{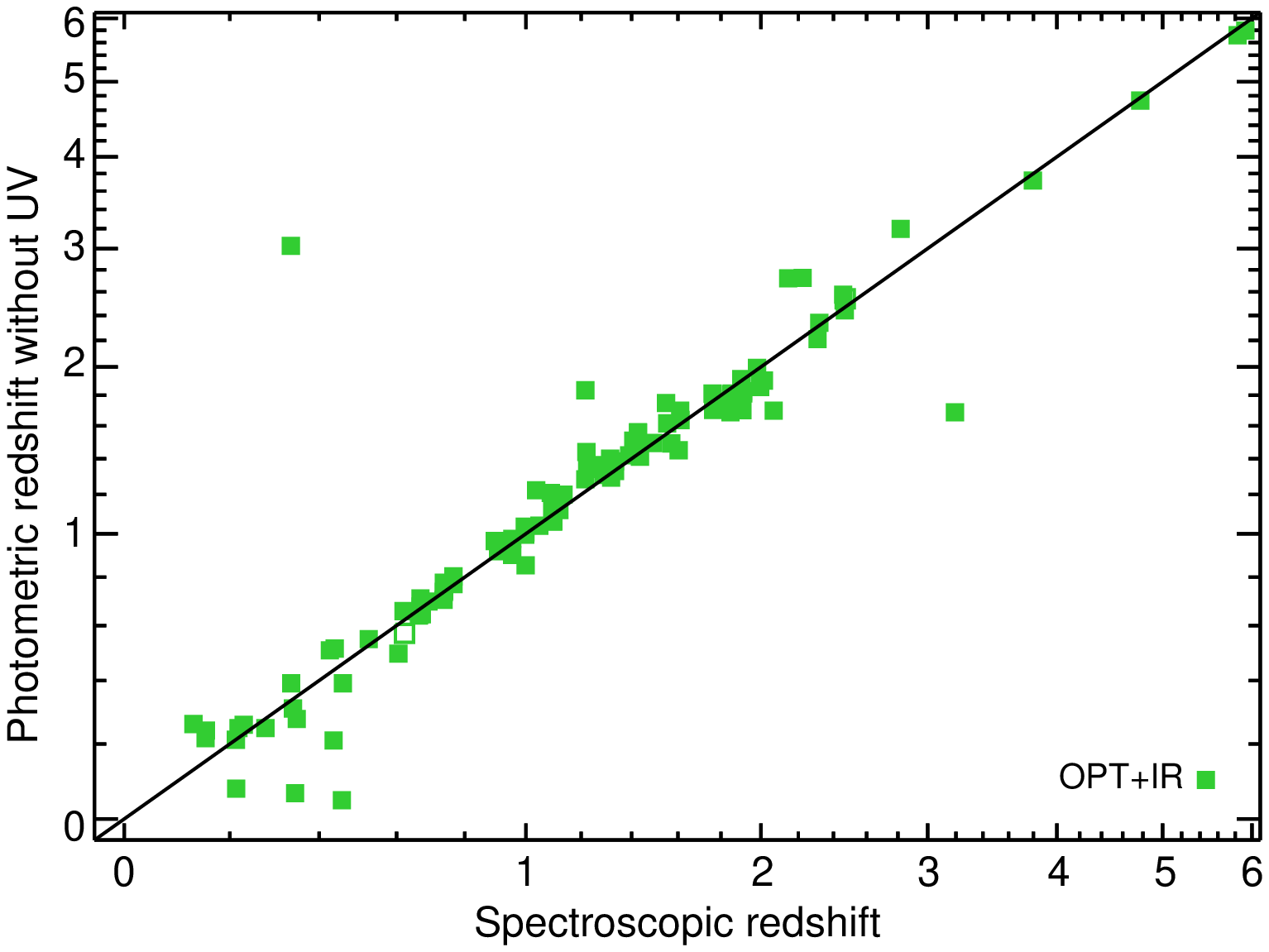}
\includegraphics[scale=0.4, viewport=5 10 550 360,clip]{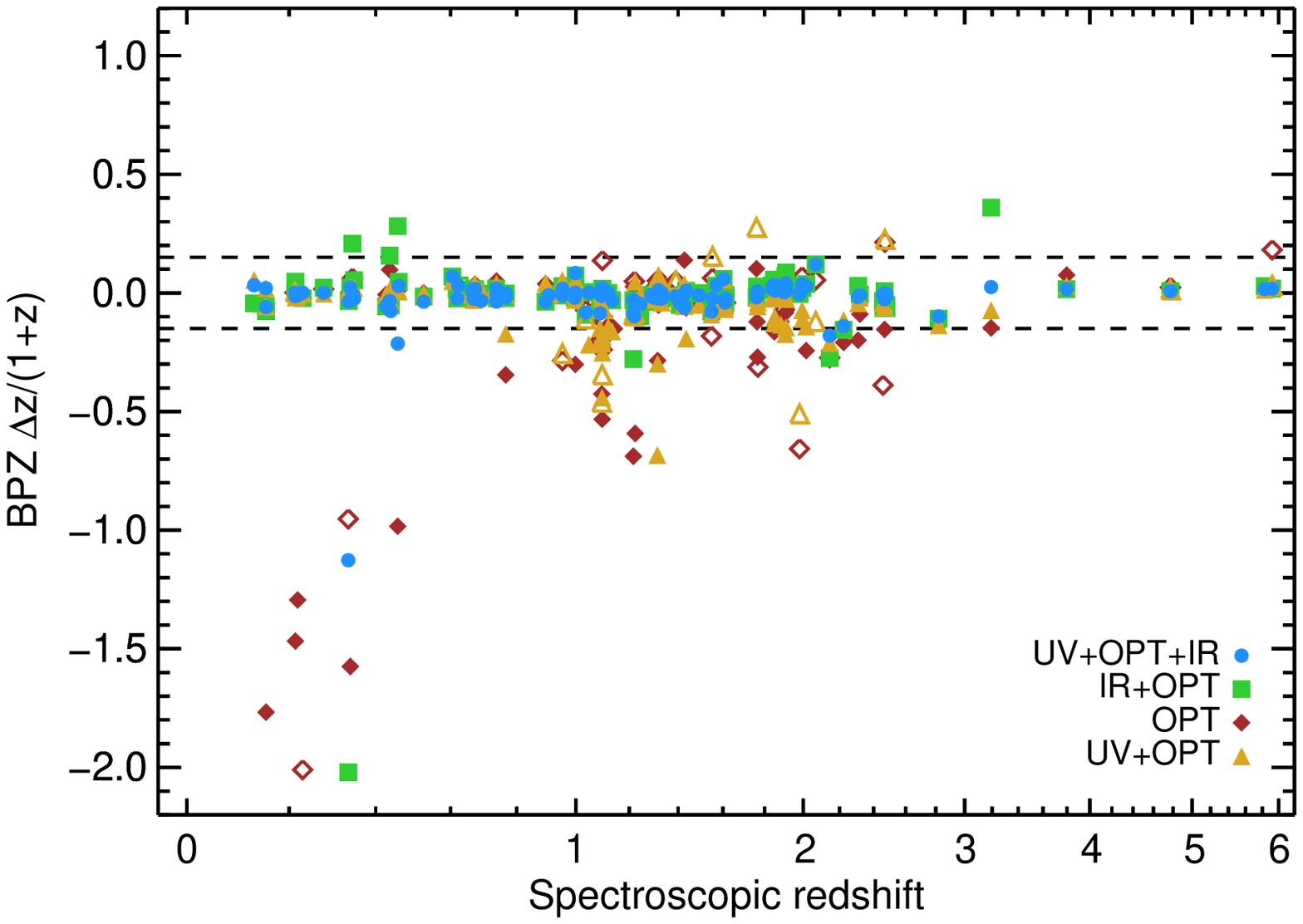}
}
\caption{ \label{fig:photoz_specz_band} 
Photometric versus spectroscopic redshifts for galaxies in the UDF on a `pseudolog' scale for the 125 galaxies with both spectroscopic coverage and NUV coverage. This shows the improved photometric redshifts with NUV and NIR data.
Each panel is for redshifts determined by including different photometric bandpasses. 
The brown diamonds are redshifts with only the optical data used, the yellow triangles are redshifts including the optical and the NUV, and the green squares are redshifts including the optical and the NIR. Filled symbols represent photometric redshifts with good {\tt ODDS} and $\chi ^2_{\mathrm{mod}}>4$, and open symbols are for redshifts that may have multiple peaks in $P(z)$ or may have a low quality fit, {\tt ODDS} $<0.9$ or $\chi ^2_{\mathrm{mod}}>4$. Dashed lines in the bottom right panel is the outlier cutoff, defined as $\left| \Delta z \right| / (1+z_{\mathrm spec}) < 0.15$. 
}
\end{figure*}
\begin{figure*}[h!]
\center{
\includegraphics[scale=0.4, viewport=5 10 550 360,clip]{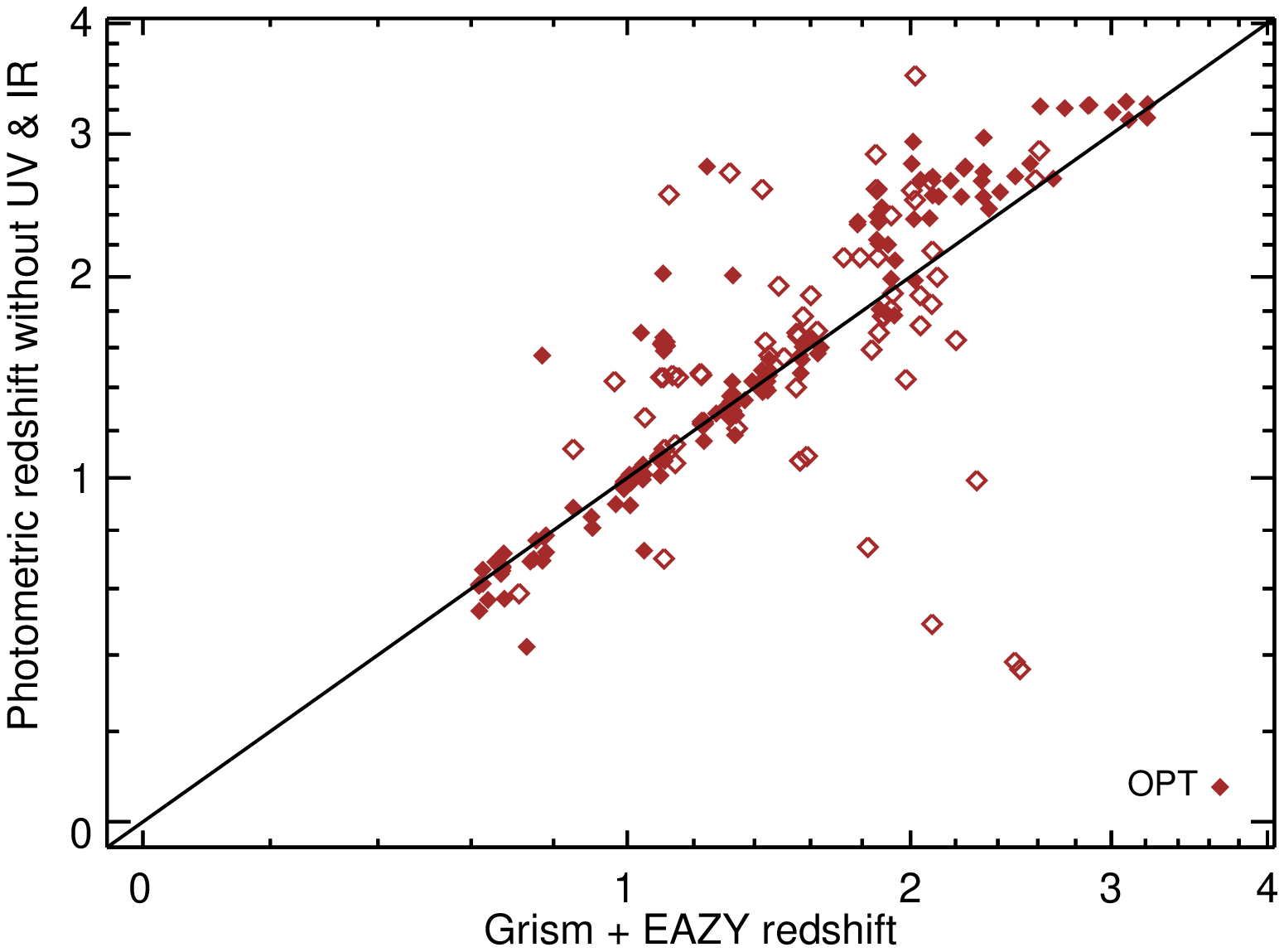}
\includegraphics[scale=0.4, viewport=5 10 550 360,clip]{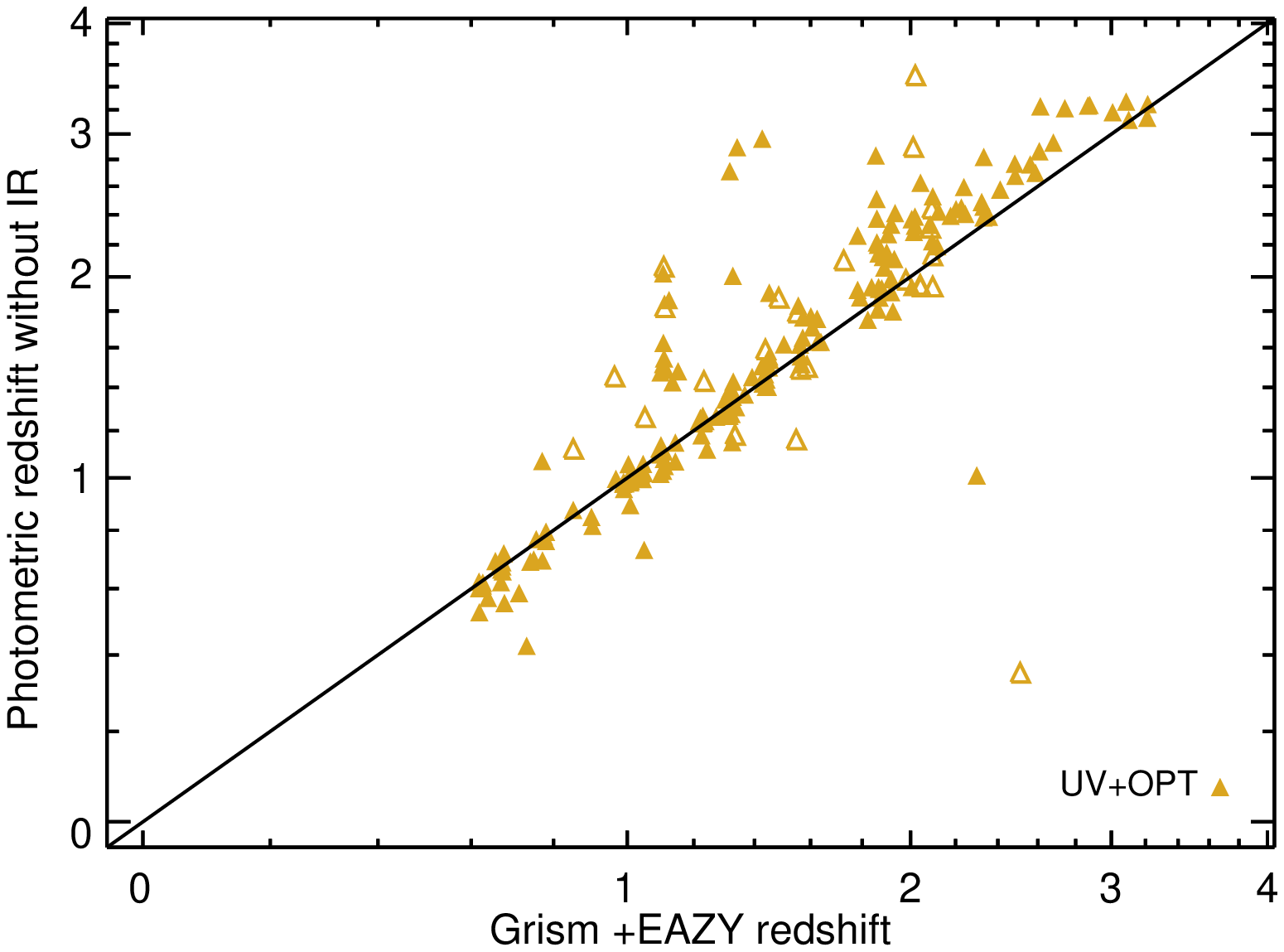}
\includegraphics[scale=0.4, viewport=5 10 550 360,clip]{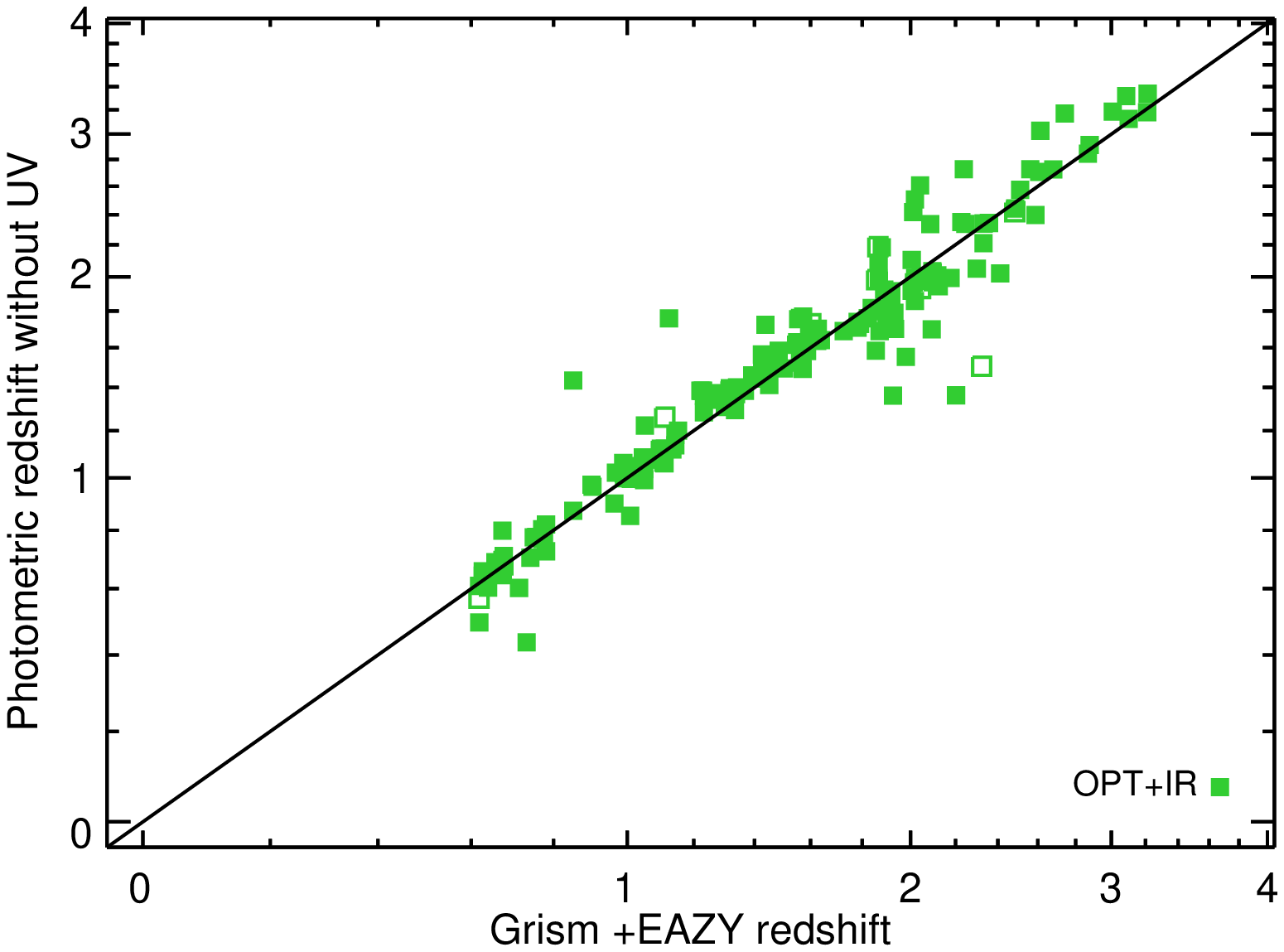}
\includegraphics[scale=0.4, viewport=5 10 550 360,clip]{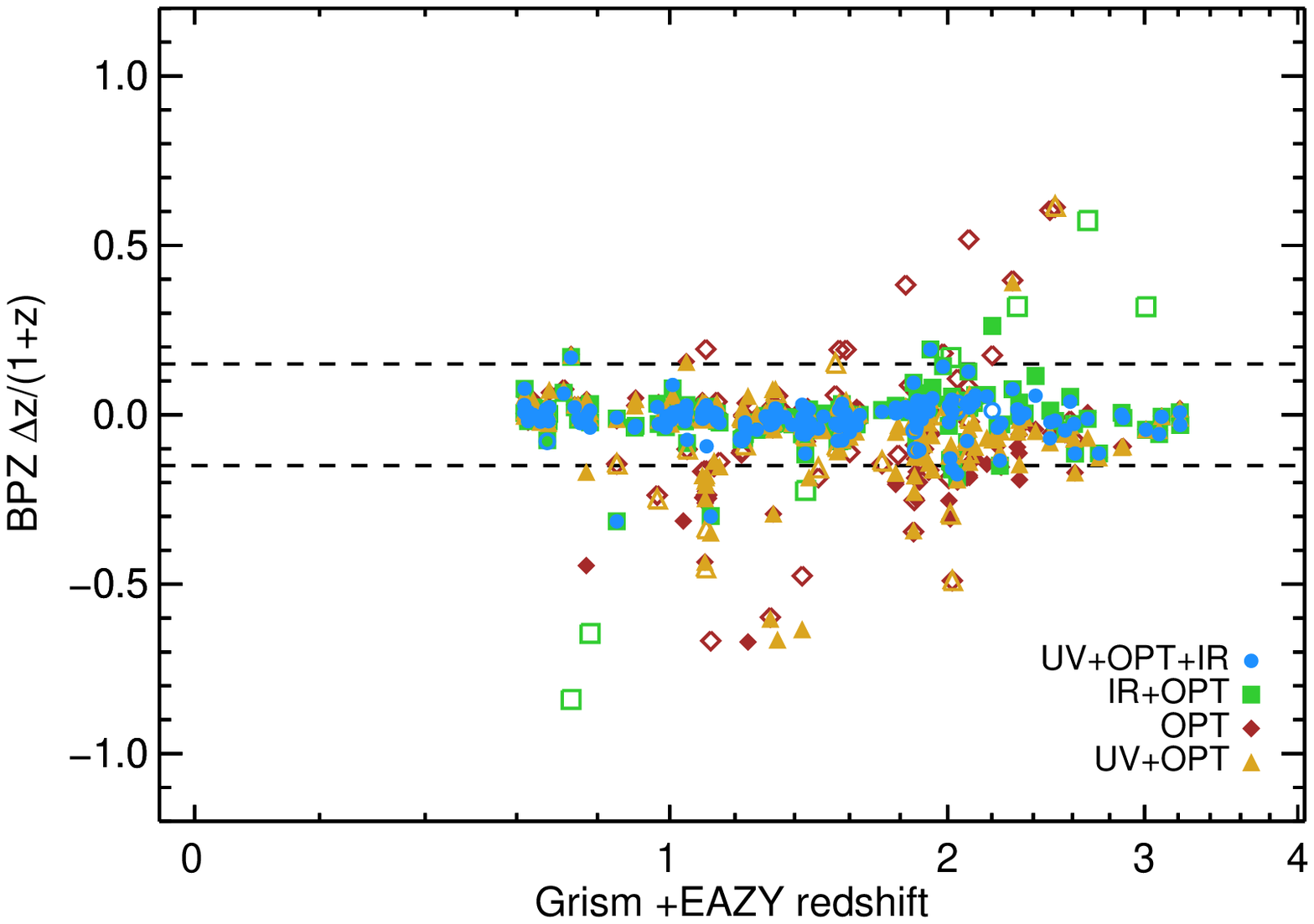}
}
\caption{ \label{fig:photoz_grismz_band} 
This figure is the same as Figure \ref{fig:photoz_specz_band}, except including the 206 grism redshifts with NUV coverage instead of the spectroscopic sample.}
\end{figure*}

\vspace{5mm}

\subsection{Improvement of Photometric Redshifts with NUV}
\label{uvimprove}
The improvements in the photometric redshifts with the addition of the NUV data can be considered by examining the spectroscopic and grism subsamples that are covered by the NUV data (125 spectroscopic redshifts, 206 grism redshifts). Before investigating this improvement, we note that the photometric redshifts presented here are already much improved over previous work due to the addition of the NIR data, even without the NUV data (similar to the redshifts determined in the deep IR section of the UDF presented in \citet{Cameron:2011}, see Section \ref{scatter}). However, the improvement of the redshifts by adding the NUV data is quantified by running BPZ on the photometry without including the NUV data, and comparing to the spectroscopic and grism redshifts. Removal of the NUV photometry results in an increase in $\sigma_{\rm NMAD}$ and OLF of the photometric redshifts. As mentioned before, for brevity only the single redshift code BPZ is considered.

\subsubsection{Spectroscopic and Grism Investigations}
For the spectroscopic redshift sample, the scatter improves  slightly, from $\sigma_{\rm NMAD}=0.029$ without NUV photometry, to $\sigma_{\rm NMAD}=0.026$ with NUV photometry.  Moreover, the OLF decreases from 6.4\% to 2.4\% (no change with good {\tt ODDS} and $\chi ^2$), which is more than a factor of 2 decrease in the OLF when including NUV data. Note that the outlier fractions are susceptible to small number statistics, with the 6.4\% corresponding to only 8 outliers. The photometric and spectroscopic redshift comparison without NUV data are shown as green squares in Figure \ref{fig:photoz_specz_band}.

For the grism redshift sample, the scatter again only improves mildly, from $\sigma_{\rm NMAD}=0.034$ without NUV photometry, to $\sigma_{\rm NMAD}=0.031$ with NUV photometry. However, the OLF decreases from 4.4\% to 2.9\% (4.0\% to 2.5\% with good {\tt ODDS} and $\chi ^2$),  which again is close to a factor of 2 decrease in the OLF. The photometric and grism redshift comparison without NUV data are shown as green squares in Figure \ref{fig:photoz_grismz_band}.

The improvement in the OLF with NUV data may be even more substantial than a factor of two, since the spectroscopic and grism redshifts preferentially sample $z<2$ (Figures \ref{fig:speczhist} and \ref{fig:grismhist}).The NUV data are expected to aid considerably at $z<0.5$ and $2<z<4$ \citep[e.g.,][]{Benitez:2000, Ilbert:2006, Rafelski:2009}, due to color redshift degeneracies resolved by the NUV data. On the other hand, the NUV data do not significantly help the scatter in redshift, likely due to the other 8 photometric bands already sufficiently constraining the redshift, other than the possibility of outliers.

In order to quantify the improvement in the redshifts with the addition of the NUV data further, the redshift determinations are compared for different combinations of bandpasses in Figures \ref{fig:photoz_specz_band} and \ref{fig:photoz_grismz_band}. The Figures show that using just the 4 optical bandpasses results in significant scatter and OLF, which is reduced by adding either NIR or NUV bandpasses. Quantitatively, for the spectroscopic sample the 4 optical bands alone yield $\sigma_{\rm NMAD}=0.072$, OLF=19.3\%, the optical with 3 NUV bandpasses yield $\sigma_{\rm NMAD}=0.071$, OLF=13.0\%, and the optical with the 4 NIR bandpasses yield $\sigma_{\rm NMAD}=0.029$, OLF=6.5\%. This is compared to $\sigma_{\rm NMAD}=0.026$, OLF=2.4\% for the spectroscopic sample including all 11 bandpasses. All outlier percentages are for good {\tt ODDS} and $\chi ^2$ on the sample that is covered by NUV data. 

Similar statistics are obtained when examining the grism redshift sample, where the optical only bandpasses yield $\sigma_{\rm NMAD}=0.064$, OLF=18.4\%, the optical and NUV bandpasses yield $\sigma_{\rm NMAD}=0.060$, OLF=14.0\%, and the optical with NIR bandpasses yield  $\sigma_{\rm NMAD}=0.034$, OLF=4.0\%. This is compared to $\sigma_{\rm NMAD}=0.031$, OLF=2.5\% for the grism sample including all 11 bandpasses.

This shows that either adding the NUV or the NIR significantly improves the redshifts, and including both yields the best results. The addition of the NIR data appears to provide a more significant improvement than the NUV overall, but not at all redshifts. For instance, the inclusion of the NUV data significantly reduces the scatter and OLF at $z<0.5$ compared to the NIR data. This is accomplished by sampling the 4000\AA~break with multiple filters, compared to only observing a flux decrement in the F435W band. 

Measurements of the Lyman break at $z>2$ should also improve the redshifts of those galaxies, although the spectroscopic sample does not include sufficient galaxies at these redshifts to show this. While the grism sample includes a larger sample at $z>2$, some fraction of the outliers observed in Figure \ref{fig:photoz_grismz_band} are possibily due to the grism redshift being incorrect, similar to the three galaxies in Figure \ref{fig:grismcheck}. Also, the good agreement of the grism + EAZY redshifts when including only the optical and NIR data are somewhat biased by the fact that the same photometry is already included in the determination of the comparison grism + EAZY redshifts. 

In addition, it is important to consider that the UDF NUV and NIR observations are not observed in an equal fashion. First, the NIR data have 4 bandpasses instead of the 3 for the NUV data. Second, the NIR data are deeper than the NUV data, consisting of 253 orbits of HST time compared to 46 for the NUV data. Third, the NIR filters are wider, have higher throughput than the NUV filters, and cover a larger wavelength range (Figure \ref{fig:filters}). Lastly, most galaxies are significantly brighter in the observed NIR than observed NUV. All together, even if the NUV data improved the photometric redshifts more than the NIR, our data would not necessarily show it. Regardless, even in the presence of deep NIR data, the NUV reduce the OLF of the sample by a factor of 2. 

The scatter and OLF for any specific redshift range may be significantly more improved than for the entire sample. The NUV data is expected to improve the color degeneracy at low and high redshifts \citep{Ilbert:2006, Rafelski:2009}, and indeed this is observed at low redshift, and insufficient spectroscopic redshifts exist at high redshift. The bottom left panel in Figure \ref{fig:photoz_specz_band} which plots redshifts determined from optical and NIR data shows quite a bit of scatter at $z<0.5$, while the top right panel for the redshifts determined with optical and NUV data shows very little scatter at those redshifts. 


\subsubsection{Comparison of Redshifts with and without NUV }

\begin{figure}[b!]
\center{
\includegraphics[scale=0.4, viewport=5 10 550 360,clip]{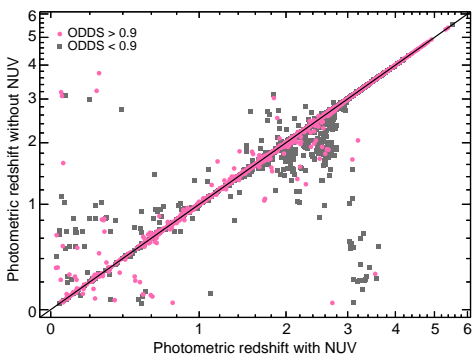}
}
\caption{ \label{fig:photozphotoz} 
Comparison of photometric redshifts with and without including the NUV data in the photometric fits for the 6459 galaxies with F606W magnitude $<29$ and with $\chi ^2_{\mathrm{mod}} <4$. The pink circles have {\tt ODDS}$>0.9$, and the gray squares have {\tt ODDS}$<0.9$. This shows that the addition of NUV data significantly affects the redshifts of galaxies in the UDF. If the redshifts with NUV data are correct, then the NUV data removes 2.2\% of the outliers with good ODDS and $\chi ^2_{\mathrm{mod}}$.
}
\end{figure}

The photometric redshifts obtained with and without NUV data are compared to each other in Figure \ref{fig:photozphotoz} to show how the NUV data alters the redshifts for the 6459 galaxies brighter than F606W magnitude of 29 and with $\chi ^2_{\mathrm{mod}} <4$. This magnitude cut is based on the depth of the NUV data and the typical colors of galaxies in the sample. Galaxies brighter than this magnitude may have their redshifts improved with the addition of the NUV data, while fainter galaxies are unlikely to show improvements. Together, the magnitude and $\chi ^2_{\mathrm{mod}}$ cuts result in a sample of good redshift fits in which the NUV data may contribute.

There are two scenarios that are evident in Figure \ref{fig:photozphotoz}. In the first, the photometric redshifts without the NUV data have {\tt ODDS}$>0.9$, and therefore a change in the photometric redshift with the addition of NUV data is not expected, as the probability function suggests only one possible redshift for that photometric fit. The second is if the redshifts have {\tt ODDS}$<0.9$, in which case the redshifts are uncertain and a change in redshift is not unexpected. 

An outlier fraction can be defined again by assuming that the photometric redshift with NUV data is correct, and requiring $\left| \Delta z_{phot} \right| / (1+z_{\mathrm spec}) < 0.15$. In this case, 7.0\% of the eight band optical and infrared redshifts would be at the wrong redshift. On the other hand, only 2.4\% do so with good {\tt ODDS}. If the redshifts determined with the NUV data are correct, then the inclusion of the NUV data removes at least 2.4\% of the outliers with good ODDS and $\chi ^2_{\mathrm{mod}}$. This is a similar to the improvement observed in the OLF for the spectroscopic and grism samples when adding NUV data, suggesting that the outlier rate is consistent with an improvement by a factor of 2 when including NUV data as suggested in the much smaller spectroscopic and grism samples.   

\begin{figure*}[]
\center{
\includegraphics[scale=0.4, viewport=5 10 550 360,clip]{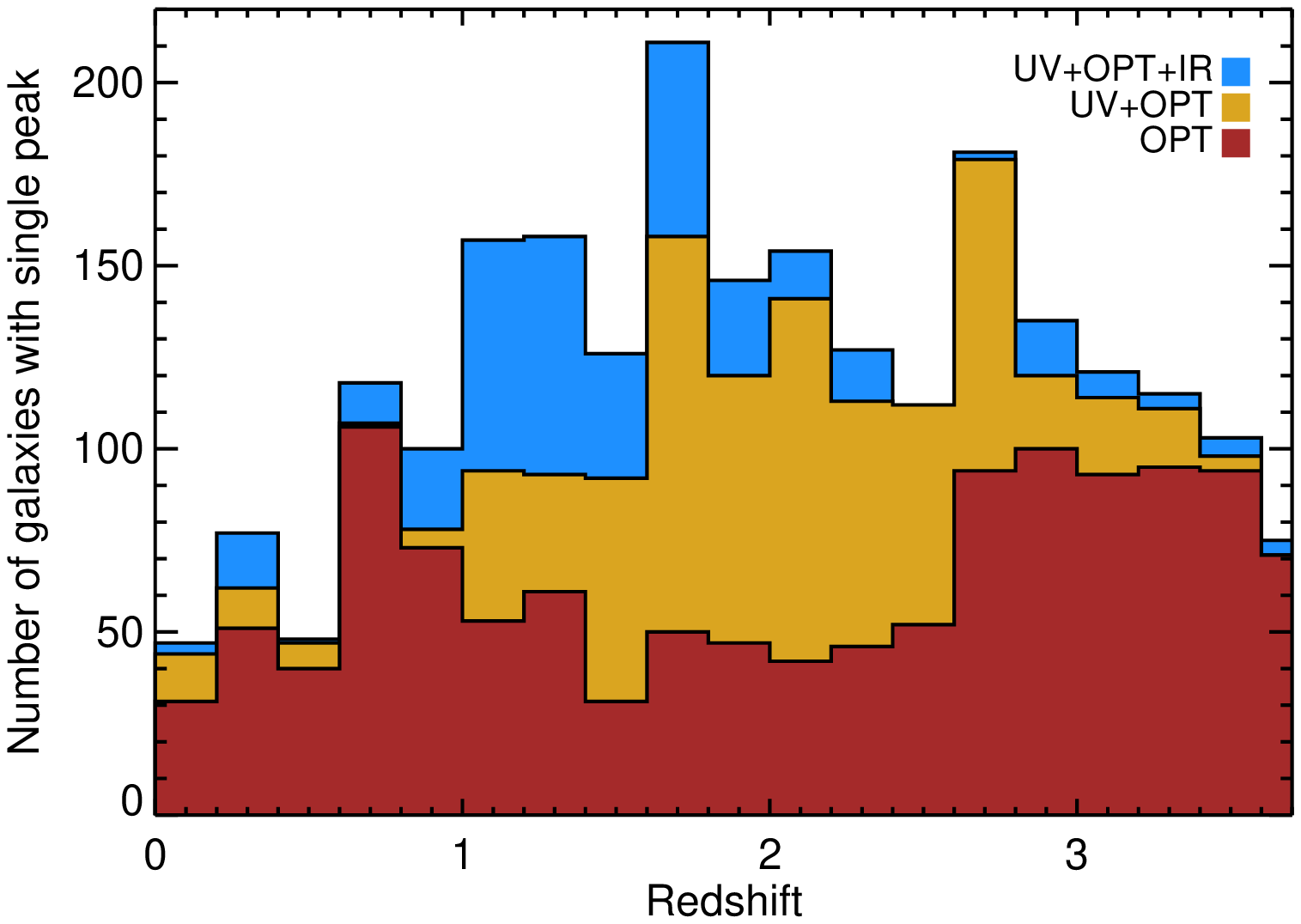}
\includegraphics[scale=0.4, viewport=5 10 550 360,clip]{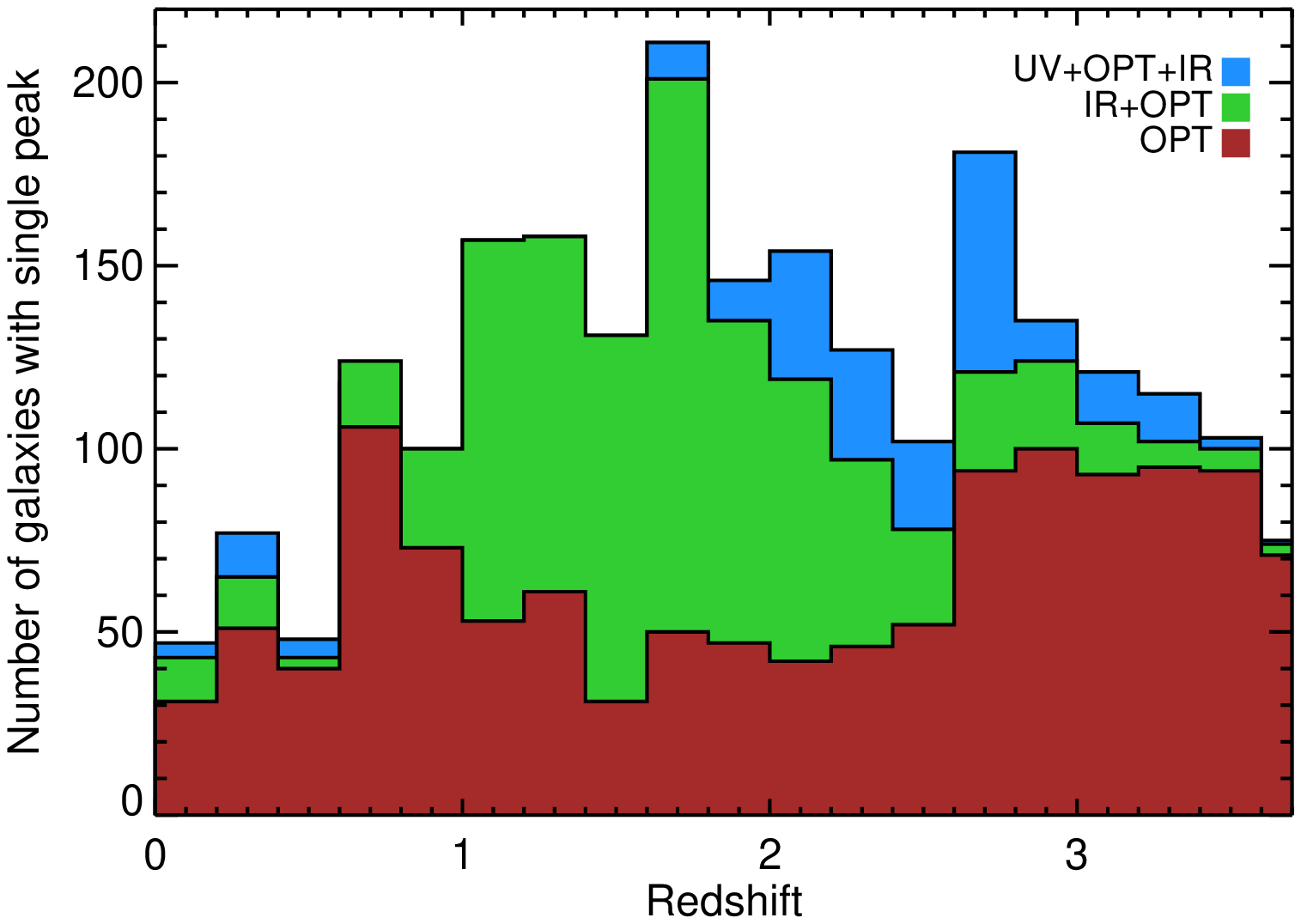}
}
\caption{ \label{fig:singlepeak} 
Number of galaxies with unambiguous photometric redshifts (a single peak in their $P(z)$) as a function of redshift for photometric redshifts including different numbers of bandpasses. This shows how different bandpasses help improve the photometric redshifts over the base redshifts determined from only optical data (brown). Photometric redshifts including only the NUV and optical bandpasses are in gold, those including only the NIR and optical are in green, and those including the NUV, optical, and NIR bandpasses are in blue. The left panel shows the improvement obtained by including the NUV and optical data (gold), and the right panel shows the same for the NIR and optical data in green. The NIR improves the redshifts more at $1\lesssim z \lesssim2$ and the NUV improves the redshifts more at $z\gtrsim2$, as the bandpasses sample the 4000\AA~ and Lyman break respectively.  
}
\end{figure*}

\subsubsection{Number of Galaxies with Single Peak in P(z)}

Another method to estimate the improvement of the photometric redshifts with the addition of the NUV data is to consider the increase in the number of galaxies with a single peak in $P(z)$ (which therefore have unambiguous redshifts) with the addition of the NUV data, as done in the predictions in \citet{Teplitz:2013}. This is similar to requiring a good ODDS value, but provides a little more information on $P(z)$, as it is possible for a redshift to have {\tt ODDS}$>0.9$ and still have a second peak in the probability distribution. 

Figure \ref{fig:singlepeak} shows the number of galaxies with a single peak as a function of redshift for different numbers of bandpasses included in the photometric fit. The photometric redshifts improve with the addition of each set of bandpasses, with the inclusion of all the bandpasses yielding the largest number of unambiguous galaxy redshifts. In the redshift range $1\lesssim z \lesssim3$ either adding the NUV or the NIR data yield similar improvements, while the addition of the NIR improves the redshifts more at $1\lesssim z \lesssim2$ and the NUV improves the redshifts more at $z\gtrsim2$. This is sensible, as the addition of the NIR data samples the 4000\AA~break at lower redshifts, and the addition of the NUV data enables sampling the Lyman break for galaxies at higher redshifts. While including both the NIR and the NUV data increases the number of single peaked galaxies the most, the improvement from adding either to the other two is not as significant as with either the addition of the NIR or NUV to just the optical.

The increase in the number of galaxies with a single peak in $P(z)$ with the addition of the NUV data to the optical and NIR data is less than the number presented in \citep{Teplitz:2013}, although the total number of galaxies is similar. There are multiple factors contributing to this difference. First, the number of galaxies predicted to have good redshifts with optical and NIR data is larger than shown in \citet{Teplitz:2013}, which results in less change with the addition of the NUV data. The cause of the lower numbers in  \citet{Teplitz:2013} is likely related to an increased number of filters and sensitivities from the UDF12 data that were not included in this original estimate \citep{Koekemoer:2013, Ellis:2013}. 

Second, the NUV data presented here are less sensitive than those used in the predictions in \citet{Teplitz:2013} because this paper only uses the half of the data which are not binned and include the addition of postflash to the images (making them more robust to CTE degradation, although also less sensitive). The other more sensitive half of the data is binned, and no pixel-based CTE corrections exist yet for binned data. Since the photometry without such corrections would be incorrect \citep[see][]{Teplitz:2013}, those data are not included here. If a pixel-based CTE correction for the binned data could be made, it would improve the redshifts further, especially for fainter galaxies. 

\section{Catalog of the UDF}
\label{catalog}

The UDF catalog is made available in both a FITS table and an ASCII table at the UVUDF website\footnote{\url{http://www.rafelski.com/uvudf/catalogs.html}} and on MAST\footnote{\url{http://archive.stsci.edu/prepds/uvudf/}}. As the table has a large number of columns, the FITS table is recommended for use. The catalog columns are described in detail in Table \ref{tab:cat}. The final catalog is trimmed to only include sources that fall within the central 11.4 arcmin$^2$ of the image based on a minimum 30\% exposure time in the optical UDF. This cut reduces the area covered by the shallow NIR data to 6.8 arcmin$^2$. However, measurements are made in the full UDF field of view for all bandpasses, enabling reliable photometry of sources near this artificial edge, but objects on the actual UDF edge are not included. In addition, the catalog is trimmed to include only sources with two photometric measurements at $>3\sigma$ confidence. While a small number spurious sources could still be present in the catalog, this removes the majority of them. This also removes high redshift sources that would only be detected in the F160W bandpass.

\begin{deluxetable*}{lll}
\tablecaption{Column Description of UDF Catalog \label{tab:cat}}
\tablehead{
\colhead{Column No.} &
\colhead{Column Title} &
\colhead{Description} }
\startdata
1 & ID & Object identification number \\
2 & COE\_ID & Object identification number from \citet{Coe:2006}\tablenotemark{a} (-99 if no match) \\
3 & RA & Right ascension (J2000 in units of decimal degrees)\\
4 & DEC & Declination (J2000 in units of decimal degrees) \\
5 & X & Image X pixel coordinate in the UDF mosaic \\
6 & Y & Image Y pixel coordinate in the UDF mosaic \\
7-17 & MAG\_* & Total AB magnitude of each filter\tablenotemark{b} \\
18-28 & MAGERR\_* & Total AB magnitude uncertainty of each filter\tablenotemark{c} \\
29-39 & FLUX\_* & Total flux of each filter in units of $\mu$Jy\tablenotemark{d} \\
40-50 & FLUXERR\_* & Total flux uncertainty of each filter in units of $\mu$Jy\tablenotemark{d} \\
51-61 & FLUX\_ISO\_* & {\tt SExtractor} isophotal Flux of each filter in units of e$^-$/s\tablenotemark{d} \\
62-72 & FLUXERR\_ISO\_* & {\tt SExtractor} isophotal Flux uncertainty of each filter in units of e$^-$/s\tablenotemark{d} \\
73 & APCOR & Aperture correction to convert from isophotal magnitude to total magnitude\tablenotemark{e} \\
74 & FWHM & {\tt SExtractor} F775W FWHM in units of pixel (1 pixel = 0.03\arcsec) \\
75 & AREAF & {\tt SExtractor} F775W isophotal area (filtered) above detection threshold in units of pixel$^2$ \\
76 & STELLARITY & {\tt SExtractor} F775W stellarity \\
77 & ELLIPTICITY & {\tt SExtractor} F775W ellipticity \\
78 & THETA & {\tt SExtractor} position angle in units of degrees \\
79 & UVUDF\_COVERAGE & 1 = covered by NUV data, 0 = not covered by NUV data \\ 
80 & UVUDF\_EDGEFLG & 1 = close to NUV edge or chip gap, 0 = not close to edge or chip gap \\
81 & Z\_BPZ & Bayesian photometric redshift (BPZ) \\
82 & ZMIN\_BPZ & BPZ 95\% lower limit to redshift \\ 
83 & ZMAX\_BPZ & BPZ 95\% upper limit to redshift \\
84 & ODDS\_BPZ & BPZ integrated $P(z)$ contained within $2*0.03*(1+z)$ \\
85 & CHISQ2\_BPZ & BPZ modified reduced chi square ($\chi ^2_{\mathrm{mod}}$) \\
86 & TEMPLATE\_BPZ & BPZ template number\tablenotemark{f} \\
87 & Z\_EAZY & EAZY photometric redshift \\
88 & ZMIN\_EAZY & EAZY 95\% lower limit to redshift \\ 
89 & ZMAX\_EAZY & EAZY 95\% upper limit to redshift \\
90 & ODDS\_EAZY & EAZY integrated $P(z)$ contained within $0.2(1 + z)$ \\
91 & CHISQ\_EAZY & EAZY reduced chi square ($\chi ^2_{\nu}$) \\
92 & NFOBS & Number of filters available for photometric redshift \\
93 & NF5SIG & Number of filters with signal to noise above 5 \\
94 & SPECZ & Spectroscopic redshift (-99 if no value) \\
95 & SPECZ\_REF & Reference for spectroscopic redshift\tablenotemark{g} \\
96 & GRISMZ & Grism + EAZY redshift from 3D-HST team \citep{Brammer:2012} (-99 if no value) \\
97 & STAR & Stars identified by GRAPES program \citep{Pirzkal:2005} (1=star) 
\enddata
\tablecomments{Column information for the UDF catalog, available both as a fits and ASCII file. Multiple entries are for all 11 bands, where the * represents each filter described in Table \ref{tab:obs}: F225W, F275W, F366W, F435W, F606W, F775W, F850LP, F105W, F125W, F140W, and F160W.
}
\tablenotetext{a}{Matched based on segmentation map overlap. (-99 if no match)}
\tablenotetext{b}{The total AB magnitude based on the Kron radius, including extinction, aperture, and PSF corrections. 99 if not detected at 1$\sigma$. -99 if not covered by a filter, or for NUV filters, also -99 if not detected in F435W image.}
\tablenotetext{c}{If not detected then equal to the 1$\sigma$ limiting AB magnitude. -99 if not covered by a filter, or for NUV filters, also -99 if not detected in F435W image.}
\tablenotetext{d}{If not detected then still provides measured flux or flux uncertainty. -99 if not covered by a filter, or for NUV filters, also -99 if not detected in F435W image.}
\tablenotetext{e}{Includes extinction and PSF correction}
\tablenotetext{f}{BPZ templates for 11 galaxies as shown in Figure \ref{fig:templates}: (1) Ell7 (2) Ell6 (3) Ell5 (4) Ell4 (5) ESO (6) Sbc (7) Scd (8) SB1 (9) SB2 (10)SB3 (11) SB11. The 9 interpolated galaxies between adjacent galaxies are represented as decimal places between the integer galaxy numbers.}
\tablenotetext{g}{(1) \citet{LeFevre:2004} (2) \citet{Szokoly:2004} (3) \citet{Mignoli:2005} (4)\citet{Daddi:2005ek} (5) \citet{Vanzella:2005, Vanzella:2006, Vanzella:2008, Vanzella:2009} (6) \citet{Popesso:2009} (7) \citet{Balestra:2010} (8) \citet{Kurk:2013} (9) \citet{Morris:2015}}
\end{deluxetable*}

The final catalog ID numbers identify which of the four {\tt SExtractor} runs each source measurement is from, as described in Section \ref{sources}. Sources from the deep-detection and normal deblending thresholds have ID numbers in the range 1-19,000, while those with shallow detection and normal deblending range from 20,000-29,000. Sources from the normal detection and low deblending have ID numbers in the range 30,000-49,000, while those from the shallow detection and low deblending range from 50,000-59,000. The corresponding ID numbers from \citet{Coe:2006} are included for backwards compatibility, matched using the segmentation maps of both this catalog and that of \citet{Coe:2006}. However, note that any single source may not be matched to an object in the \citet{Coe:2006} catalog, and similarly, two different sources may be matched to the same \citet{Coe:2006} source. 

The photometric magnitude and flux values are provided for each filter. The total magnitudes are aperture-matched PSF corrected total measurements, including corrections for Galactic extinction. These total measurements are based on the isophotal color and the total flux measured via the Kron aperture flux (see Section \ref{aptpsf}). The aperture correction to convert from the isophotal magnitude to total magnitude is available in the catalog, along with the original isophotal flux values. The NUV photometric measurements and resulting photometric redshifts all are based on the smaller F435W apertures as described in Section \ref{bband}. The table also indicates whether a source is covered by the NUV data, and if so, if it falls near the NUV edge or chip gap. The optical and NIR data cover the entire UDF FOV, and therefore no flag is provided. In addition, some basic morphological measurements from the F775W {\tt SExtractor} runs are also provided, such as FWHM and Ellipticity. 

The photometric redshifts from both BPZ and EAZY are presented, along with the number of filters available to the codes, and the number of measurements above 5$\sigma$ included in the fits. The quality of the photometric redshifts are included, described by {\tt ODDS},  $\chi ^2_{\mathrm{mod}}$, and $\chi ^2_\nu$ as described in Section \ref{photoz}. Care should be taken when using these redshifts to select galaxies with the appropriate values; to obtain a robust sample, we recommend using a minimum {\tt ODDS} of 0.9, maximum $\chi ^2_{\mathrm{mod}}$ of 4 for BPZ redshifts, and a maximum $\chi ^2_\nu$ of 10 for EAZY redshifts. Stricter cuts may be used for a higher confidence photometric redshift, at the cost of a smaller sample size. Spectroscopic and Grism redshifts are also provided if available, and known stars are marked.

\section{Summary}
\label{summary}

New NUV and NIR imaging obtained with WFC3 on HST provides the opportunity to significantly improve the photometric redshifts of galaxies in the UDF. The NUV data come from the UVUDF imaging campaign including three filters, F225W, F275W, and F336W \citep{Teplitz:2013}, and  cover $\sim60\%$ of the UDF FOV.  The NIR data is obtained from a combination of the UDF09, UDF12, and CANDELs data in order to obtain the deepest NIR coverage and cover the entire FOV of the UDF \citep{Oesch:2010a,Oesch:2010b,Bouwens:2011b, Grogin:2011,Koekemoer:2011, Koekemoer:2013, Ellis:2013}. 

The NUV data are newly calibrated as described in Section \ref{obs} and the Appendix, and made available to the public through MAST. In short, the data are corrected for CTE degradation, and calibrated with custom dark files, as the current STScI CDBS (CRDS) dark calibration files are no longer sufficient to calibrate the NUV data largely due to CTE degradation of the detector. The new custom dark files are CTE corrected, mostly remove a gradient and blotchy pattern from the science data, and properly flag hot pixels. The custom processing significantly improves the final mosaic image quality. Darks processed in a similar fashion should be used for any observations with low backgrounds, and STSCI plans to release such darks sometime in 2015. In order to minimize the blotchy pattern in future data, we recommend that observations should use dither patterns larger than $\sim20$ pixels to reduce the pattern in the final mosaics.

Sources are detected in an eight-band-averaged image including the optical and NIR. Aperture-matched PSF corrected photometry is performed on the eleven photometric bandpasses, and are included in Table \ref{tab:cat}. Numerous {\tt SExtractor} runs per filter were used to optimize the detection and segmentation definitions of sources in the UDF. Specifically, two different detection thresholds and two different deblending thresholds were run, and then merged together. In addition, the NUV data use separate aperture definitions determined from the F435W image to maximize SN, and minimize systematics from any leftover blotchy pattern in the NUV data (See section \ref{blotchyscience}).

Photometric redshifts were derived using two different redshift codes, BPZ \citep{Benitez:2000} and EAZY\citep{Brammer:2008} using different galaxy templates and priors. These redshifts are compared to a new compilation of 169 reliable spectroscopic redshifts of galaxies (Table \ref{tab:specz}). This comparison reveals a low scatter of  $\sigma_{\rm NMAD}=$ 0.028 for BPZ and $\sigma_{\rm NMAD}=$ 0.030 for EAZY, and low outlier fractions (OLF) of 2.4\% and 3.8\% for BPZ and EAZY respectively for good {\tt ODDS} and $\chi ^2$. Results from the comparison to grism redshifts are similar, with $\sigma_{\rm NMAD}=$ 0.031 and $\sigma_{\rm NMAD}=$ 0.040 for BPZ and EAZY and OLF of 2.5\% and 4.9\% respectively for good {\tt ODDS} and $\chi ^2$. This is an improvement of $\sim2$ in $\sigma_{\rm NMAD}$ and a factor of $\sim3$ in OLF over the last comprehensive UDF redshift catalog by \citet{Coe:2006}. 

We showed that adding the NUV data to the photometric redshift derivations in addition to the optical and NIR gave a mild improvement in $\sigma_{\rm NMAD}$, and a factor of $\sim2$ improvement in the OLF.  The improvement of the redshifts with adding NUV or NIR data to the optical data depend on the redshift, with the NIR improving the redshifts at $1\lesssim z \lesssim2$ more, and the NUV doing so at $z<0.5$ and $z\gtrsim2$. In addition, the NUV data appear to significantly reduce the scatter and OLF at $z<0.5$. It is important to consider the caveat that in these comparisons, the NUV data only consist of 46 orbits over 3 filters, versus the 253 orbits of NIR data over 4 filters. In other words, for a smaller investment of NUV observing time, the outlier fraction can be significantly improved with the addition of the NUV. Such improvements in the photometric redshifts are observed even when considering only the NUV and optical data, showing the power of including the NUV data.

The photometry and photometric redshifts of all the sources in the full UDF are presented in Table \ref{tab:cat} as both a FITS and ASCII table.  Overall, this catalog provides photometry in a new wavelength regime, and significantly improves the photometric redshifts. It will aide future galaxy research in the UDF, and has already contributed to a number of studies using the NUV data and redshifts \citep[e.g.,][]{Bond:2014, Kurczynski:2014, Mei:2014vg} and others in preparation.

\acknowledgments

 We would like to thank Sylvia Baggett and Jay Anderson at the Space Telescope Science
 Institute for their help with solving new calibration and CTE challenges in the Epoch 3 NUV data.  
 We thank Gabe Brammer, Kate Whitaker, Chun Ly, and Daniel Angerhausen for useful discussions.
We also thank the referee, Dr. Michael Drinkwater, for his helpful comments.
 Support for HST Program GO-12534 was provided by NASA
 through grants from the Space Telescope Science Institute, which is
 operated by the Association of Universities for Research in
 Astronomy, Inc., under NASA contract NAS5-26555.
MR also acknowledges support from an appointment to the NASA Postdoctoral Program at Goddard Space Flight Center,
as well as support from HST GO-13389 and HST GO-13309.
 \facility{} Facilities:  HST (WFC/ACS, WFC3/UVIS, WFC3/IR)

\begin{appendices}
\section{WFC3/UVIS Dark Calibrations}

The UVUDF data analysis revealed a number of improvements that can be made in the creation of the dark calibration files. We worked closely with STScI to solve problems with the darks and communicate the best practices for future observing strategy and data reduction. The need for improved darks are caused by changes in the characteristics of the detector (such as CTE degradation) since the methodology for dark creation for WFC3/UVIS was developed. There are three effects discussed here: hot pixel detection, background gradients, and a blotchy background pattern. The official dark calibrations are part of the Calibration Reference Data System (CDBS), and the STScI released darks will hereafter be referred to as CDBS darks.  We note that recently STScI switched over to the Calibration Reference Data System (CRDS), which currently uses the same CDBS darks.

Part of the WFC3/UVIS calibration process is the subtraction of a dark reference file to remove the dark current and to identify hot pixels that can cause significant artifacts in the images. STScI releases new darks every 3-5 days, which is necessary due to the appearance of $\sim500$ new hot pixels per day. The increase in hot pixels is mitigated by annealing the detector once per month, removing $\gtrsim70$\% of the hot pixels \citep{Borders:2009}. Even so, the number of permanent hot pixels which are not removed by anneals is increasing by 0.05-1\% per month (WFC3 instrument handbook). 

The current methodology used for the CDBS darks is to combine the raw darks obtained over a 3-5 day period to create superdarks for each time period, composed of $\sim10-20$ dark exposures with integration times of $\sim900$s each. The process is described in detail in \citet{Martel:2008} and \citet{Borders:2009}, but the general process is outlined here. First, the process cleans the raw darks of cosmic rays, and then creates an average of all the raw dark files. A threshold is then used to find all pixels deviant from the median dark value. These pixels are marked as hot, and then the final superdark is the median value of the average dark (a single value for all good pixels), with the hot pixels superimposed and marked in the data quality array. This is done to minimize introducing noise in the data from the uncertainty in the dark current per pixel from a small number of exposures over the 3-5 day period. While this process was sufficient for the data products shortly after installation on HST, the detector characteristics have changed making this process insufficient for current WFC3/UVIS images. 

\subsection{Missed Hot Pixels}
\label{hotpix}
The first issue with the CDBS darks is that an outdated definition of a hot pixel is applied \citep[originally determined in][]{Borders:2009}, which results in unmasked warm-to-hot pixels remaining in data obtained in recent years. Under this definition, pixels with values $>0.015$e$^-$/s are flagged as hot. However, since the time that this threshold was determined the characteristics of the detector have changed, especially due to CTE degradation. The situation is worse because the CDBS darks are not CTE corrected, resulting in hot pixels being missed even if they would otherwise have been above the threshold level, and are present in the CTE corrected science data. The leftover hot pixels that remain unmasked yield significant artifacts for mosaics calibrated with the CDBS darks. 

\begin{figure}[]
\center{
\includegraphics[scale=0.4, viewport=10 10 550 360,clip]{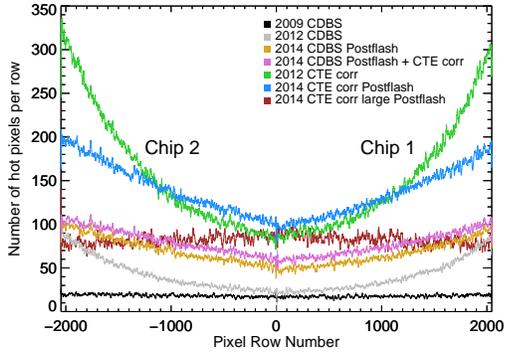}
}
\caption{ \label{fig:hotpix} 
The number of hot pixels per row versus row number for WFC3/UVIS data determined from dark data. The center of the two CCDs is at pixel row number 0 at the center of the plot, and the edges of the chips near the readouts are at the right and left side of the figure. The black, gray, and yellow lines are determined from CDBS darks from 2009, 2012, and 2014 respectively, with the 2014 dark based on raw darks that are post-flashed. The green and blue lines include  the more aggressive hot pixel masking from the new dark methodology described in Section \ref{newdarks} for the same 2012 and 2014 darks as the previous ones. The pink, green, blue, and brown lines also include an additional CTE correction step on the raw dark files, before the creation of a superdark. The brown line is for a 2014 superdark obtained with high level of postflash to minimize CTE trails. As expected, the decrement of the number of hot pixels in rows near the center of the chip is not present in this dataset. This figure shows that the CDBS darks from 2012 and 2014 are missing a substantial number of hot pixels, especially for rows far from the readout. The yellow and blue lines also shows the improvement in hot pixel detection in the post-flash data.  }
\end{figure}

These missed hot pixels are more prevalent far from the readout, where CTE degradation plays a more important role. This effect is evident in Figure \ref{fig:hotpix}, which shows the number of hot pixels as a function of row number, where the center of the two chips is at pixel row 0, and the readout of the two CCDs are at the left and right sides of the plot. The black line shows the number of hot pixels per row in the first in-flight CDBS dark released in 2009, and a constant number of hot pixels per row is observed. All lines on this plot after this are expected to have an increased number of hot pixels with time, due to the growth of the number of permanent hot pixels  \citep{Borders:2009}. However, the gray line shows the hot pixels for one of the dark files associated with the UVUDF data in 2012. At the edges of the plot, the number of hot pixels is indeed higher than the black line, but at the center of the plot far from the readouts, the number of hot pixels per row decreases significantly. 

There is no reason to expect that the number of hot pixels is physically developing differently on different parts of the chips. This effect is due to the CTE degradation causing hot pixels to be missed far from the readout. Therefore, a significant number of hot pixels are not masked in the CDBS darks at later times, as the gray line would be expected to be flat from one side of Figure \ref{fig:hotpix} to the other. Therefore, the 2012 CDBS darks are missing 57\% of the hot pixels deemed important to mask by \citep{Borders:2009}. 

The missed hot pixel issue is somewhat reduced by the introduction of post-flash darks in October 2012, and the hot pixels per row in a dark from 2014 is shown in yellow in Figure \ref{fig:hotpix}. However, even in the post flash data, 35\% of the hot pixels are missed. If these data are CTE corrected before being processed in otherwise the same manner as the CDBS darks (pink line), then 25\% of the hot pixels are missed. This shows that even if a CTE correction is applied to the post-flashed raw darks, there are still a lot of hot pixels missed. Since the CTE correction code aims to not over correct the pixels, it is likely that these hot pixels are missed due to imperfect CTE corrections. The fact that the post flash darks (yellow and pink curves) are flatter than the darks without post flash (gray curve) in Figure \ref{fig:hotpix} is evidence that a large fraction of the missed pixels are due to CTE degradation. 

\subsection{Background Gradient}

The second issue with the CDBS darks is that the median value of the average darks is applied as the value of all pixels in the dark frame. This median dark file is not suitable for low background data, because it leaves a low-level gradient. This gradient is typically small compared to the sky background in exposures with high sky backgrounds, however, in the low-background UV imaging, it is the dominant structure. The use of CDBS darks therefore results in a significant gradient in the science mosaics.

\begin{figure}[]
\center{
\includegraphics[scale=0.4, viewport=5 10 550 360,clip]{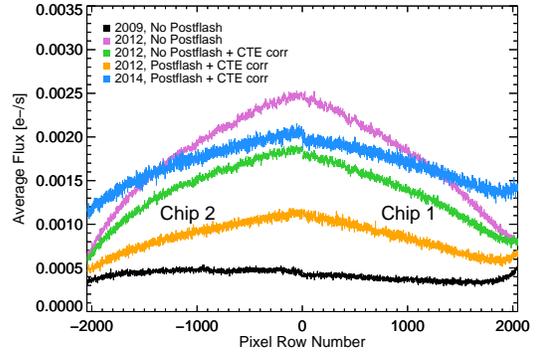}
}
\caption{ \label{fig:gradient} 
The average background flux per row versus row number for WFC3/UVIS darks. The center of the two CCDs is at pixel row number 0 at the center of the plot, and the edges of the chips near the readouts are at the right and left side of the figure. The black, red, green and blue lines are from darks obtained in 2009, 2012, and 2014. The green, blue, and orange lines also include a CTE correction. The orange line represents the dark background used for calibrating the post-flashed UVUDF data based on early test postflashed dark exposures obtained in 2012 as described in Section \ref{newdarks}. The CTE correction and post-flashing the raw darks both significantly reduce the background gradient, but does not remove it.
}
\end{figure}

The background gradients measured in the averaged darks for different time periods are shown in Figure \ref{fig:gradient}. Similar to Figure  \ref{fig:hotpix}, the center of the two CCDs is at pixel row number 0 at the center of the plot, and the edges of the chips near the readouts are at the right and left side of the figure. The black line shows the first in-flight dark background gradient, and it is relatively flat. This explains why the CDBS darks do not attempt to correct this background gradient. However, the darks associated with the UVUDF in 2012 have a significant gradient, shown as the pink line in Figure \ref{fig:gradient}. The background level is significantly higher at the center of the chips far from the readout than at the edges close to the readout, with a factor of four variation in the background level from the center to the edge of the chip. This suggests that CTE degradation may be responsible for the observed gradient, with flux from cosmic rays and hot pixels smeared out into the background. 

The green line in the plot shows the gradient from the same data as the pink line, but with a CTE correction applied. This dark shows a reduction in the background gradient, confirming that at least part of the gradient from the pink line is indeed due to CTE degradation. In addition, the blue line shows a dark obtained in 2014 with post-flash enabled and with a CTE correction applied. Although the gradient is further reduced, it still has a factor of two variation across the chip. 

The most likely cause is that the CTE degradation strategies and corrections do not fully correct the data for the CTE degradation, although there could be other physical causes for the remaining gradient in the background. One such possibility is if the postflash light is not fully subtracted, as the postflash illumination is not flat, but instead is $\sim30$\% brighter near the center of the chip \citep{MacKenty:2012}. A slight imperfection in the subtraction of the postflash could contribute to the $\sim1$e$^-$ gradient left over in the dark observations. Since the science data are flashed at the same level as the darks, any such light would be removed when the darks are subtracted and thus could be missed in tests.

\subsection{Blotchy Background Pattern}
\label{blotchy}
\begin{figure}[]
\center{
\includegraphics[scale=0.4]{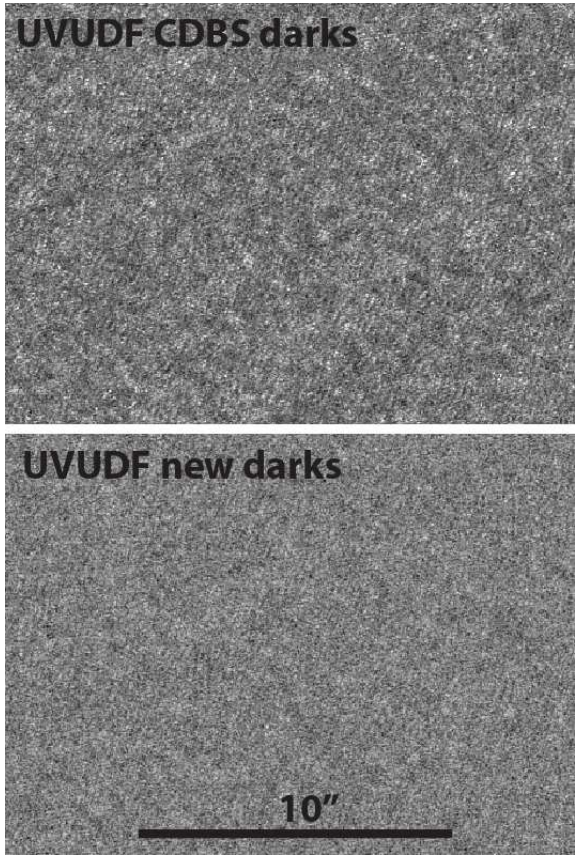}
}
\caption{ \label{fig:blotchy} 
Both panels show the same empty region in the UDF in the F275W filter with WFC3/UVIS. The top panel is processed with the CDBS darks released by STScI, and the bottom panel with the new improved darks described in Section \ref{newdarks}. The top panel exhibits a blotchy pattern which is significantly reduced in the bottom panel after applying the improved darks to the science data. }
\end{figure}

In addition to the background gradient, the dark background exhibits a blotchy pattern on scales of $\sim20$ pixels. This pattern is difficult to characterize, as the signal to noise (SN) in each 3-5 day dark is very low. In addition, the darks contain many hot pixels and artifacts, which make the noise pattern difficult to visualize in the darks themselves. Therefore, the pattern is characterized and shown in the science data in an empty region of the UDF. The top panel in Figure \ref{fig:blotchy} shows an empty region of the UDF with a non-uniform background noise pattern. This background pattern exhibits blotches that are $\sim0.001-0.002$ e$^-$/s in magnitude (both positive and negative) with $\sim20$ pixel diameters ($\sim0.8$ arcseconds). This translates to changes of $\sim1-2$ e$^-$ in the original raw dark frames. 
Given this blotchy pattern in the data, future observations should use dither patterns larger than $\sim20$ pixels to reduce the pattern in the final mosaics.

\subsection{New Dark Methodology}
\label{newdarks}
Due to these limitations of the CDBS darks, and the need for cleaner image mosaics in the deep UDF observations, 
we developed a new methodology to create improved darks for WFC3/UVIS superdarks that mitigate the issues described above. 
The new dark methodology begins by applying the pixel-based CTE correction on all the individual raw dark files. Those data are then used in two different (but connected) steps; the first is to determine the hot pixels, and the second is to produce the best background level. For the first step, the darks from a 3-5 day window are used to create a superdark in a somewhat similar manner as done by STScI in creating the CDBS darks. The same time frame as the CDBS darks is employed, and thus the data are on the same cadence as CDBS. A short cadence is necessary for masking the hot pixels, since they vary rapidly. The hot pixels are determined in a different fashion than the CDBS darks. For each 3-5 day dark, the background per averaged dark is modeled with a 3rd order polynomial, which is used to remove the gradient. After the gradient is removed, hot pixels are found with a more aggressive threshold. This step is used to define the hot pixels in the darks on short time scales, and also provides cosmic ray cleaned raw darks for the later step.

\subsubsection{Finding Hot Pixels}

The threshold for finding hot pixels is set based on the darks in 2012 and has been optimized for the UDF observations.  A different threshold is used for the post-flashed and non-postflashed darks. The thresholds are set such that the number of hot pixels per row at the center of the chip is equal to the number of hot pixels in the original CDBS darks close to the readout. In this fashion, all hot pixels that would have been originally masked are found.  This results in a 3.7$\sigma$ threshold for non-postflashed data, and a 4.9$\sigma$ threshold for post-flash data. The standard deviation of the pixels is determined from an iterative 3$\sigma$ rejection, thereby removing any hot pixels from the distribution before $\sigma$ is determined. Also, the background gradient is removed before applying the threshold, such that the gradient does not cause different numbers of hot pixels to be found across the chip. We ran tests with different thresholds to determine if a higher or lower threshold would produce better science images, and found that thresholds similar to those determined by STScI produced the right balance of masking all the noticeable hot pixels, while not masking too much of each raw data frame. 

The resulting number of hot pixels as a function of pixel row number for our new darks are shown in Figure \ref{fig:hotpix}. The green line shows the resultant hot pixel numbers for a CTE corrected dark from 2012, and the blue line shows the same for a 2014 CTE corrected post-flashed dark. The number of pixels at the center (pixel row 0) in the green and blue lines match the number of hot pixels per row  near the readout for the equivalent CDBS dark as shown in the gray and yellow lines respectively, confirming the threshold levels. In order to mask the hot pixels far from the readout, a higher fraction of pixels are masked than actual hot pixels in the data.

As a test of the CTE degradation affecting hot pixels, a recent set of calibration darks obtained by STScI in March 2014 were post-flashed to a very high level, $\sim90$e$^-$. This enables a check on the cause of the variation in number of hot pixels per row across the chip, as darks with a very large postflash count are expected to experience fewer charge traps, and therefore less CTE loss. The brown line in Figure \ref{fig:hotpix} is from a superdark processed using the methodology described above, and is mostly flat, which is consistent with the expectation that when CTE losses are minimized, hot pixels are not preferentially lost far from the readout. However, the postflash subtraction with such a large postflash is not perfect, and the darks contain residual postflash light. Therefore, investigations of the background gradient is not possible with this dataset, nor can one apply the original CDBS threshold for a hot pixel. The slight increase in the number of hot pixels at the center of the chip is likely due to the $\sim30$\% variation in the illumination pattern of the postflash, changing the noise properties at the center of the chip where the illumination is brightest. 

One downside of more aggressive hot pixel flagging is that a lot of pixels are flagged as hot that are likely just warm pixels, and do not need to be flagged as hot. This is especially evident in Figure \ref{fig:hotpix} for rows near the readout, where a factor of more than two times the number of hot pixels are flagged than the rows furthest away from the readout. However, the fraction of unnecessarily flagged pixels of the detector is small, at $\sim1$\% and $\sim2$\% of the chips for data with and without post-flash respectively. Even with this increase, the total fraction of hot pixels flagged remains small, at $\sim3$\% and $\sim4$\% for with and without post-flash respectively. On the other hand, if the same thresholds are applied to the initial inflight 2009 darks (black line), the number of hot pixels would be doubled. Regardless, this increased hot pixel flagging should not be of major consequence, since the total number of hot pixels remains small, and flagging the hot pixels improves image quality sufficiently to justify this choice.

\subsubsection{Removing Gradient and Blotchy Pattern}

The second step in the new dark methodology is to obtain the best dark background to handle the gradient and blotchy pattern described above. Specifically, we create a dark with spatial structure in it, rather than a dark with a single value for all good pixels. To do so, the CTE corrected cosmic ray cleaned images from a single anneal cycle are averaged together, with the hot pixels from each 3-5 day window being masked. This determines the actual dark level for each good pixel. This averaged dark per anneal cycle is then used in conjunction with the hot pixel map on the 3-5 day period to create a new superdark consisting of the average dark over the entire anneal cycle, with hot pixels masked at the shorter cadence. 
We determined that using a shorter time period than an entire anneal cycle did not improve the blotchy pattern, and the characteristics of the darks changed sufficiently between anneals to preclude averaging together raw darks from different anneal cycles. Also, note that the modeled dark gradients in the first step are not used other than to find the hot pixels. 

The average dark per anneal intrinsically includes any gradient in the dark background, as well as the observed blotchy pattern.  The resulting darks significantly improve the image quality in the UVUDF, as well as in other WFC3/UVIS programs. The improvement is evident in Figure \ref{fig:blotchy}, where the top panel shows an empty region of the NUV UDF data processed with the CDBS darks, and the bottom panel shows the same data processed with the new dark calibrations. While the blotchy pattern is not removed completely, it is significantly reduced. 

In order to obtain the best hot pixel mask and dark background, it is best to match post-flashed science data with post-flashed darks, and CTE corrected science data with CTE corrected darks. Not doing so can yield missed hot pixels in the science frames, and imperfect dark background levels. The UVUDF data were obtained before STScI started obtaining post-flashed darks on a regular basis. However, 30 dark frames with post-flash were obtained during the testing process in the same month as the UVUDF science images, and are the final dark exposures used to calibrate the UVUDF.

A comparison of the science mosaics shows that the blotchy pattern is less prominent when the science data are processed with darks based on these 30 darks, rather than the higher SN darks produced from a larger number of non-postflashed darks obtained at the same time. Since no post-flash darks exist on the same days as the UVUDF science exposures, we include the hot pixel masks from the non-postflashed darks in addition to the hot pixel mask from the post-flashed darks in the darks for the UVUDF. In this way more hot pixels are flagged than are real, but this ensures that most of the hot pixels are masked. The final hot pixel mask is therefore equivalent to the green line in Figure \ref{fig:hotpix}, and the average background flux is shown as the orange line in Figure \ref{fig:gradient}. 

\subsection{Impact of Darks On Science Photometry}
\label{blotchyscience}

\begin{figure}[b!]
\center{
\includegraphics[scale=0.4, viewport=10 10 550 360,clip]{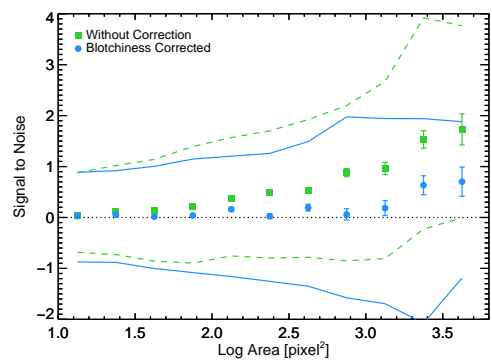}
}
\caption{ \label{fig:snarea} 
Median values of the distribution of measured F275W SN measurements and their standard deviation as a function of source aperture area. The green squares and dashed line show this  for science mosaics without a blotchiness correction, and the blue circles and solid line show it when including a blotchiness correction. If the sky noise distribution were uncorrelated and Gaussian, the distribution of the SN would have a median of zero and a standard deviation equal to one, independent of aperture area. However, an increase in both the median and standard deviation of the SN values is observed as the aperture area increases. The application of the improved darks significantly improves the photometry in the NUV, bringing the median closer to zero, and the standard deviation closer to one. }
\end{figure}

In order to test the effects of the blotchiness on the catalog photometry, an empty source catalog was generated in which the source apertures computed from the detection image were placed on blank regions in the UV images.  Figure \ref{fig:snarea} shows the distribution of the measured NUV SN for these blank sources as a function of aperture area.  If the sky noise distribution were uncorrelated and Gaussian, the distribution of the SN for blank sources would have a median of zero and a standard deviation equal to one, independent of aperture area. However, an increase in both the median and standard deviation of the SN values is observed as the aperture area increases, suggesting that a {\tt SExtractor} run on the UVUDF images would yield measurements with excess flux and underestimated uncertainties for large apertures.

This flux excess and uncertainty underestimation is caused by the blotchiness pattern described in Section \ref{blotchy}.  The number of positive $3\sigma$ outliers in the distribution of flux values in a blank region of one of the UVUDF images is $\sim 4$ times that of the negative $3\sigma$ outliers. Since {\tt SExtractor} computes the sky background using a $3\sigma$ clipped mean, these outliers will not be accounted for in the sky subtraction and will be falsely attributed to flux from the source. In addition, the blotchy pattern introduces a noise pattern that is not accounted for in the RMS images used to measured the uncertainty of sources. 

This effect is significantly stronger for science mosaics using darks without a correction for the blotchiness, compared to mosaics using darks that do. The green squares and dashed line in Figure \ref{fig:snarea} correspond to measurements on science data using uncorrected darks, and the blue circles and solid line represents the same for darks corrected for the blotchiness. Specifically, the uncorrected darks are equivalent to the CDBS darks, except that they include a correction for the background gradient and the hot pixels. The blotchiness corrected darks are created as described in Section \ref{newdarks}, and include the new average dark background per anneal cycle, which removes much of the blotchy pattern. After correcting the science mosaics for the blotchiness via the darks, the median S/N drops from $\sim 0.9$ to $\sim 0.1$ for 1000-pixel apertures.  Even after the correction for blotchiness, there are still some residual systematic effects at very large
apertures, so photometry in that regime should be approached with caution. No corrections are applied in the catalogs for the remaining blotchy pattern, as only a small number of sources have such large areas, and the blotchiness corrected images behave sufficiently well for the majority of the sources in the catalog.

\end{appendices}

\bibliography{uvudf,manual}

\end{document}